\documentclass[a4paper]{article}

\usepackage{amsmath,amssymb,amsfonts,graphicx,latexsym}
\usepackage[english]{babel}
\usepackage{xcolor}
\usepackage{natbib}
\bibpunct{[}{]}{;}{n}{}{,} 
\usepackage{graphics}
\date{}
\usepackage[margin=25mm]{geometry}

\newcommand{\ie}{{\em i.e.} }
\newcommand{\eg}{{\em e.g.} }
\newcommand{\ca}{{\mathcal A}}
\newcommand{\cb}{{\mathcal B}}

\newcommand{\cd}{{\mathcal D}}

\newcommand{\cm}{{\mathcal M}}
\newcommand{\cn}{{\mathcal N}}

\newcommand{\ct}{{\mathcal T}}
\newcommand{\cu}{{\mathcal U}}

\newcommand{\cx}{{\mathcal X}}
\newcommand{\cy}{{\mathcal Y}}

\newcommand{\ua}{\ve{\alpha}}
\newcommand{\ub}{\ve{\beta}}

\newcommand{\Nn}{{\mathbb N}}

\newcommand{\Rr}{{\mathbb R}}

\newcommand{\ve}[1]{\boldsymbol{#1}}			
\newcommand{\Ve}[1]{\boldsymbol{#1}}			

\newcommand{\acc}[1]{\left\{#1\right\}}
\newcommand{\enu}{ , \, \dots \,,}

\newcommand{\prt}[1]{\left(#1\right)}			
\newcommand{\abs}[1]{\left| #1 \right|}			

\newcommand{\vare} {\varepsilon}
\newcommand{\Var}[1]{{\rm Var}\left[ #1 \right]}

\newcommand{\Esp}[1]{{\mathbb E}\left[ #1 \right]}

\newcommand{\eqdef}{\stackrel{\text{def}}{=}}
\newcommand{\tr}{^{\textsf T}}				
\newcommand{\dsp}[1]{\displaystyle{#1}}

\newcommand{\R}{\mathbb{R}}
\newcommand{\ff}{\mathbf{f}}
\newcommand{\bbeta}{\boldsymbol{\beta}}
\newcommand{\ttheta}{\boldsymbol{\theta}}

\newcommand{\FF}{\mathbf{F}}
\newcommand{\X}{\mathbf{X}}
\newcommand{\GP}[2]{\mathrm{GP}\left(#1, #2\right)}
\newcommand{\rr}{\mathbf{r}}
\newcommand{\K}{\mathbf{R}}
\newcommand{\E}[1]{\mathbb{E}\left[ #1 \right]}
\newcommand{\V}[1]{\mathrm{Var}\left( #1 \right)}
\newcommand{\EZ}[1]{\mathbb{E}_Z\left[ #1 \right]}
\newcommand{\VZ}[1]{\mathrm{Var}_Z\left( #1 \right)}
\newcommand{\CovZ}[2]{\mathrm{Cov}_Z\left( #1, #2 \right)}

\newcommand{\imx}{\boldsymbol{\mathsf{A}}}
\renewcommand{\cm}{G}
\newcommand{\mpci}{G^{\textsf{PC}\backslash i}} 
\newcommand{\mpc}{G^{\textsf{PC}}}

\usepackage{authblk}
	
\usepackage[english]{babel}
\graphicspath{{./},{./figures/}}

\title{Metamodel-based sensitivity analysis: Polynomial chaos
	expansions and Gaussian processes}
\author[1]{Lo\"ic Le Gratiet}
\author[2]{Stefano Marelli}
\author[2]{Bruno Sudret}

\affil[1]{EDF R\&D, 6 quai Watier, 78401 Chatou, France} 

\affil[2]{ ETH Z\"urich, Chair of Risk, Safety \& Uncertainty
  Quantification, Stefano-Franscini-Platz 5, CH-8093 Z\"urich,
  Switzerland}

\begin{document}

\maketitle

\begin{abstract}
  Global sensitivity analysis is now established as a powerful approach
  for determining the key random input parameters that drive the
  uncertainty of model output predictions. Yet the classical computation
  of the so-called Sobol' indices is based on Monte Carlo simulation,
  which is not affordable when computationally expensive models are
  used, as it is the case in most applications in engineering and
  applied sciences. In this respect metamodels such as polynomial chaos
  expansions (PCE) and Gaussian processes (GP) have received tremendous
  attention in the last few years, as they allow one to replace the
  original, taxing model by a surrogate which is built from an
  experimental design of limited size. Then the surrogate can be used to
  compute the sensitivity indices in negligible time. In this chapter an
  introduction to each technique is given, with an emphasis on their
  strengths and limitations in the context of global sensitivity
  analysis. In particular, Sobol' (resp. total Sobol') indices can be
  computed analytically from the PCE coefficients. In contrast,
  confidence intervals on sensitivity indices can be derived
  straightforwardly from the properties of GPs. The performance of the
  two techniques is finally compared on three well-known analytical
  benchmarks (Ishigami, G-Sobol and Morris functions) as well as on a
  realistic engineering application (deflection of a truss structure).\\

\end{abstract}

\section{Introduction}

In modern engineering sciences computational models are used to simulate
and predict the behavior of complex systems. The 
governing equations of
the system are usually discretized so as to be solved by dedicated
algorithms. In the end a computational model (a.k.a. simulator) is built
up, which can be considered as a mapping from the space of input
parameters to the space of quantities of interest that are computed by
the model. However, in many situations the values of the parameters
describing the properties of the system, its environment and the various
initial and boundary conditions are not perfectly well-known. To account for 
such uncertainty, they 
are typically described by possible variation ranges or probability
distribution functions.

In this context global sensitivity analysis aims at determining which input
parameters of the model influence the most the predictions, \ie how the
variability of the model response is affected by the uncertainty of the
various input parameters. A popular technique is based on the
decomposition of the response variance as a sum of contributions that
can be associated to each single input parameter or to combinations
thereof, leading to the computation of the so-called Sobol' indices.

As presented earlier in this book (see Variance-based 
sensitivity analysis: Theory and estimation algorithms), 
the use of Monte Carlo simulation to compute  Sobol' 
indices requires a large number of samples (typically,
thousands to hundreds of thousands), which may be an 
impossible requirement when the underlying computational 
model is expensive-to-evaluate. 
To bypass this
difficulty, {\em surrogate models} may be built. Generally speaking, a
surrogate model (a.k.a. metamodel or emulator) is an approximation of
the original computational model:
\begin{equation}
  \label{eq:BST-001}
 \ve{x} \in \cd _{X} \subset \Rr^d \mapsto  y =\cm(\ve{x}) 
\end{equation}
which is constructed based on a limited number of runs of the true
model, the so-called {\em experimental design}:
\begin{equation}
  \label{eq:BST-002}
  \cx= \acc{\ve{x}^{(1)} \enu \ve{x}^{(n)}}.
\end{equation}
Once a type of surrogate model is selected, the parameters 
have to be fitted based on the information contained in the 
experimental design $\cx$ and associated runs of the 
original computational model $\cy =
\acc{y_i = G(\ve{x}^{(i)}), \, i=1 \enu n}$. Then the 
accuracy of the surrogate shall be estimated by some kind 
of validation technique. For a general introduction to 
surrogate modelling the reader is referred to
\cite{storlie2009implementation} and to the recent review 
by \citet{IoossReview2015}.

In this chapter we discuss two classes of surrogate models that are
commonly used for sensitivity analysis, namely {\em polynomial chaos
  expansions} (PCE) and {\em Gaussian processes} (GP). The use of
polynomial chaos expansions in the context of sensitivity analysis has
been originally presented in \citet{SudretCSM2006, SudretRESS2008b}
using a non intrusive least-square method. Other 
non-intrusive strategies for the calculation of PCE 
coefficients include spectral projection through sparse 
grids (\eg\citet{Crestaux2009, 
Buzzard2011,Buzzard2012}) 
and sparse polynomial expansions (\eg
\citet{BlatmanRESS2010}). 
In the last five years numerous application examples have 
been developed using PCE for sensitivity analysis, \eg
\citet{Fajraoui2011,Younes2013,Brown2013,Sandoval2012}. Recent
extensions to problems with dependent input parameters can be found in
\citet{Sudret2013a,Munoz2013}.

In parallel, Gaussian process modeling has been 
introduced in the
context of sensitivity analysis by \citet{Welch1992,Oak04, Marrel2008,
  Marrel2009}. Recent developments in which metamodeling errors are
taken into account in the analysis have been proposed by
\citet{MultiFidelitySA,ChastaingLeGratiet2015}.

The chapter first recalls
the basics of the two approaches and details how they can be used to
compute sensitivity indices. The two approaches are then compared on
different benchmark examples as well as on an application in structural
mechanics.

\section{Polynomial chaos expansions}
\label{sec:7-02}

\subsection{Mathematical setup}
\label{sec:7-02.1}
Let us consider a computational model $\cm:\ve{x} \in \cd_{\Ve{X} }
\subset \Rr^d \mapsto y = \cm(\ve{x}) \in \Rr$. Suppose that the uncertainty
in the input parameters is modeled by a random 
vector $\Ve{X} $ with
prescribed joint probability density function (PDF) $f_{\Ve{X}}(\Ve{x})$. The
resulting (random) quantity of interest $Y= \cm(\Ve{X})$ is obtained by
propagating the uncertainty in $\Ve{X} $ through $\cm$.  Assuming that
$Y$ has a finite variance (which is a physically meaningful assumption
when dealing with physical systems), it belongs to the so-called Hilbert
space of second order random variables, which allows for the following
spectral representation to hold \citep{Soize2004}:
\begin{equation}
  \label{eq:700}
  Y = \sum_{j=0}^\infty y_{j}\, Z_j.
\end{equation}
The random variable $Y$ is therefore cast as an infinite
series, in which $\acc{Z_j}_{j=0}^\infty$ is a numerable set of random
variables (which form a basis of the Hilbert space), and
$\acc{y_j}_{j=0}^\infty$ are coefficients. The latter may be interpreted
as the {\em coordinates} of $Y$ in this basis. In the sequel we focus on
{\em polynomial chaos expansions}, in which the basis terms
$\acc{Z_j}_{j=0}^\infty$ are multivariate orthonormal polynomials in the
input vector $\Ve{X}$, \ie $Z_j = \Psi_j(\Ve{X} )$.

\subsection{Polynomial chaos basis}
\label{sec:7-02.2}
In the sequel we assume that the input variables are statistically {\em
  independent}, so that the joint PDF is the product of the $d$ marginal
distributions: $f_{\Ve{X}}(\ve{x}) = \prod_{i=1}^d f_{X_i}(x_i)$, where the
$f_{X_i}(x_i)$ are the marginal distributions of each variable $\acc{X_i, \, i=1
  \enu d}$ defined on $\cd_{X_i}$.  For each single variable $X_i$ and
any two functions $\phi_1, \phi_2:x\in \cd_{X_i} \mapsto \Rr$, we define
the functional inner product by the following integral (provided it
exists):
\begin{equation}
  \label{eq:701}
  \langle \phi_1, \phi_2 \rangle_{i} \eqdef   \Esp{\phi_1(X_i)\,
    \phi_2(X_i)}  = \int_{\cd_{X_i}} \phi_1(x) \,
  \phi_2(x) \, f_{X_i}(x) \, dx.
\end{equation}
Using the above notation, classical algebra allows one to build a family
of {\em orthogonal polynomials} $\{ P_k^{(i)},\:k \in \mathbb{N}\} $
satisfying
\begin{equation}
  \left\langle   {P_j^{(i)},P_k^{(i)}} \right\rangle_{i} 
  \eqdef
  \Esp{P_j^{(i)} (X_i)\, P_k^{(i)}(X_i)} =
  \int_{\cd_{X_i}} 
  P_j^{(i)}(x)\;P_k^{(i)}(x)\;{f_{{X_i}}}(x)\,dx =  a_j^{(i)}\,\delta _{jk},
\label{eq:702}
\end{equation}
see \eg \citet{Abramowitz}. In the above equation subscript~$k$ denotes
the degree of the polynomial $P_k^{(i)}$, $\delta_{jk}$ is the Kronecker
symbol equal to 1 when $j=k$ and 0 otherwise and 
$a_j^{(i)}$ corresponds to
the squared norm of $P_j^{(i)}$:
\begin{equation}
  \label{eq:703}
  a_j^{(i)} \eqdef \parallel P_j^{(i)}\parallel^2_i \; = \left\langle
    {P_j^{(i)},P_j^{(i)}} \right\rangle_{i} .
\end{equation}
In general orthogonal bases may be obtained by applying the Gram-Schmidt
orthogonalization procedure, \eg to the canonical family of
monomials $\acc{1,\, x,\, x^2 ,\dots}$. For standard 
distributions, the associated families of orthogonal 
polynomials are well-known \citep{Xiu2002}.
For instance, if $X_i \sim \cu(-1,1)$ has a uniform distribution over
$[-1,1]$, the resulting family is that of the so-called {\em Legendre
polynomials}. When $X_i \sim \cn(0,1)$ has a standard normal distribution 
with zero mean value and unit standard deviation, the
resulting family is that of the {\em Hermite polynomials}. 
The families
associated to standard distributions are summarized in
Table~\ref{tab:701} (taken from \citet{SudretHDR}).

\begin{table}[!ht]
  \centering\caption{Classical families of orthogonal polynomials (taken
    from \citet{SudretHDR})}
  \label{tab:701}
  \begin{tabular}[c]{lllc}
    \hline\\[-0.5em]
    Type of variable & Distribution  & Orthogonal polynomials &
    Hilbertian basis $\psi_k(x)$ \\[0.3em]\hline\\[-0.3em]
    \begin{tabular}[c]{c}
    Uniform \\
    $\cu(-1,1)$\\[0.5em]
    \end{tabular}
    & ${\mathbf 1}_{[-1,1]}(x) /2 $  & Legendre $P_k(x)$ &
    $P_k(x)/  \sqrt{\frac{1}{2k+1}}$ \\
    \begin{tabular}[c]{c}
      Gaussian \\
      $\cn(0,1)$\\[0.5em]
    \end{tabular}
 & $ \frac{1}{\sqrt{2 \pi}} e^{-x^2/2}$ & Hermite
    $H_{e_k}(x)$ &  $H_{e_k}(x)/  \sqrt{k!}$\\
    \begin{tabular}[c]{c}
      Gamma \\
      $\Gamma(a,\lambda=1)$\\[0.5em]
    \end{tabular}& $x^a\, e^{- x} \,{\mathbf 1}_{\Rr^+} (x)$ &
    Laguerre $L^a_k(x)$ & $L^a_k(x)/\sqrt{\frac{\Gamma(k+a+1)}{k!}}$\\
    \begin{tabular}[c]{c}
      Beta \\
      $\cb(a,b)$\\[0.5em]
    \end{tabular}
    & ${\mathbf 1}_{[-1,1]}(x) \, \frac{(1-x)^a(1+x)^b}{B(a)\,
      B(b)} $ & Jacobi $J^{a,b}_k(x)$ 
    &  $J^{a,b}_k(x)/{\mathfrak{J}}_{a,b,k}$ \\
    & & \multicolumn{2}{c}{
      $\mathfrak{J}_{a,b,k}^2 
      = \frac{2^{a+b+1}}{2k+a+b+1}
      \frac{\Gamma(k+a+1)\Gamma(k+b+1)}{\Gamma(k+a+b+1) \Gamma(k+1)}$}  \\
    \hline
  \end{tabular}
\end{table}

\noindent Note that the obtained family is usually not orthonormal. By enforcing
normalization, an {\em orthonormal family} $\acc{ \psi_j^{(i)}
}_{j=0}^\infty$ is obtained from Eqs.(\ref{eq:702}),(\ref{eq:703}) as follows
(see Table~\ref{tab:701}):
\begin{equation}
  \label{eq:704}
  \psi_j^{(i)} = P_j^{(i)} / \sqrt{a_j^{(i)}} \quad i=1\enu d , \quad j\in
  \Nn.
\end{equation}
From the sets of univariate orthonormal polynomials one can now build
{\em multivariate} orthonormal polynomials with a {\em tensor product}
construction. For this purpose let us define the multi-indices $\ua \in
\Nn^d$, which are ordered lists of integers:
\begin{equation}
  \label{eq:705}
  \ua = \prt{\alpha_1 \enu    \alpha_d} \,, \quad \alpha_i \in \Nn.
\end{equation}
One can associate a multivariate polynomial $\Psi_{\ua}$ to any
multi-index $\ua$ by
\begin{equation}
  \label{eq:706}
  \Psi_{\ua}(\ve{x}) \eqdef \prod_{i=1}^d \psi_{\alpha_i}^{(i)} (x_i),
\end{equation}
where the univariate polynomials $\acc{\psi_k^{(i)}, \, k\in \Nn}$ are
defined above, see Eqs.(\ref{eq:702}),(\ref{eq:704}). By virtue of
Eq.(\ref{eq:702}) and the above tensor product construction, the
multivariate polynomials in the input vector $\Ve{X} $ are also
orthonormal, \ie
\begin{equation}
  \label{eq:707}
  \Esp{\Psi_{\ua} (\Ve{X} )\, \Psi_{\ub} (\Ve{X} )} \eqdef
  \int_{\cd_{\Ve{X}}} \Psi_{\ua} (\Ve{x} )  \Psi_{\ub} (\Ve{x} ) \,
  f_{\Ve{X} }(\ve{x}) \, d \ve{x}  =
  \delta_{\ua \ub} \qquad \forall \, 
  \ua, \ub \in \Nn^d,
\end{equation}
where $ \delta_{\ua \ub}$ is the Kronecker symbol which is equal to 1 if
$\ua = \ub$ and zero otherwise. With this notation, it can be proven
that the set of all multivariate polynomials in the input random vector
$\Ve{X} $ forms a basis of the Hilbert space in which $Y=\cm(\Ve{X})$ is
to be represented \citep{Soize2004}:
\begin{equation}
  \label{eq:708}
  Y =\sum_{\ua \in \Nn^d}  y_{\ua} \, \Psi_{\ua}(\Ve{X}). 
\end{equation}

\subsection{Non standard variables and truncation scheme}
\label{sec:7-02.3}
In practical sensitivity analysis problems the input variables may not
necessarily have standardized distributions as the ones described in
Table~\ref{tab:701}. Thus {\em reduced variables} $\ve{U}$ with standardized 
distributions are introduced first through an isoprobabilistic 
transform:
\begin{equation}
  \label{eq:709}
  \Ve{X} = \ct(\Ve{U}).
\end{equation}
For instance, when dealing with independent uniform distributions with
support $\cd_{X_i} = [a_i, b_i], \, i=1\enu d$, the
isoprobabilistic transform reads:
\begin{equation}
  \label{eq:710}
  X_ i = \frac{a_i +b_i}{2} +   \frac{b_i - a_i }{2}\, U_i
 \qquad U_i \sim\cu([-1,1]) .
\end{equation}
In the case of Gaussian independent variables $\acc{X_i \sim \cn(\mu_i \,,\,
  \sigma_i)\,, \; i=1 \enu d}$, the one-to-one mapping reads:
\begin{equation}
  \label{eq:711}
  X_ i =  \mu_i +\sigma_i\  U_i, \qquad U_i \sim\cn(0,1) 
\end{equation}
In the general case when the input variables are non Gaussian (\eg Gumbel 
distributions, see application in Section~\ref{sec:7-04.3}), the one-to-one 
mapping may be obtained as follows:
\begin{equation}
  \label{eq:712}
  X_ i =   F_{X_i}^{-1}\prt{\Phi(U_i)} \qquad U_i \sim\cn(0,1) 
\end{equation}
where $F_{X_i}$ (resp. $\Phi$) is the cumulative distribution function
(CDF) of variable $X_i$ (resp. the standard normal CDF).

This isoprobabilistic transform approach also allows one to address problems
with {\em dependent} variables. For instance, if the input vector $\Ve{X} $
is defined by a set of marginal distributions and a Gaussian copula,
it can be transformed into a set of independent standard normal
variables using the Nataf transform \citep{Ditlevsen1996,Lebrun2009a}.

The representation of the random response in Eq.(\ref{eq:708}) is exact
when the infinite series is considered. However, in practice, only a
finite number of terms may be computed. For this purpose a {\em
  truncation scheme} has to be selected. Since the polynomial chaos
basis consists of multivariate polynomials, it is natural to consider as a 
truncated series all the polynomials up to a given maximum degree. Let us 
define the {\em total degree} of a multivariate polynomial $\Psi_{\ua}$ by:
\begin{equation}
  \label{eq:712b}
  \abs{\ua}  \eqdef \sum_{i=1}^d \alpha_i.  
\end{equation}
The {\em standard truncation scheme} consists in selecting all
polynomials such that $\abs{\ua}$ is smaller than or equal to a given
$p$. This leads to a set of polynomials denoted by $ \ca^{d,p} =
\acc{\ua \in \Nn^d \; : \; \abs{\ua} \le p}$ of cardinality:
\begin{equation}
  \label{eq:712c}
  \text{card}~ \ca^{d,p}= \binom{d+p}{p} = \frac{(d+p)!}{d!  \, p!}.
\end{equation} 

The maximal polynomial degree $p$ may typically be equal to $3-5$ in
practical applications. Note that the cardinality of $ \ca^{d,p}$
increases exponentially with $d$ and $p$. Thus the number of terms in
the series, \ie the number of coefficients to be computed, increases
dramatically when $d$ is large, say $d>10$. This complexity is referred to as 
the {\em curse of dimensionality}. This issue may be solved using specific
algorithms to compute sparse PCE, see \eg
\citet{SudretJCP2011,Doostan2011}.

\subsection{Computation of the coefficients and error estimation}
\label{sec:7-02.4}
The use of polynomial chaos expansions has emerged in the 
late eighties in uncertainty quantification problems under 
the form of {\em stochastic   finite element methods} 
\citep{Ghanembook1991}. In this setup the
constitutive equations of the physical problem are 
discretized both in the physical space (using standard 
finite element techniques) and in the random space using 
polynomial chaos expansion. This results in coupled systems 
of equations which require ad-hoc solvers, thus the term 
``intrusive approach''. 

{\em Non intrusive} techniques such as projection or stochastic
collocation have emerged in the last decade as a means to compute the
coefficients of PC expansions from repeated evaluations of the existing
model $\cm$ considered as a black-box function.  In this section we
focus on a particular non intrusive approach based on least-square
analysis.

Following \citet{BerveillerPMC04,Berveiller2006a}, the computation of
the PCE coefficients may be cast as a least-square minimization problem
(originally termed ``regression'' problem) as follows: once a truncation
scheme $\ca \subset \Nn^d$ is chosen (for instance, $\ca = \ca^{d,p}$),
the infinite series is recast as the sum of the truncated series and a
residual:
\begin{equation}
  \label{eq:713}
  Y = \cm(\Ve{X} ) = \sum_{\ua \in \ca} y_{\ua} \, \Psi_{\ua}(\Ve{X} ) + \vare,
\end{equation}
in which $\vare$ corresponds to all those PC polynomials whose index
$\ua$ is not in the truncation set $\ca$. The least-square minimization
approach consists in finding the set of coefficients $\ve{y}=
\acc{y_{\ua}, \, \ua \in \ca}$ which minimizes the mean square error
\begin{equation}
  \label{eq:714}
  \Esp{\vare^2} \eqdef \Esp{\prt{\cm(\Ve{X})-\sum_{\ua \in \ca} y_{\ua}
      \, \Psi_{\ua}(\Ve{X})}^2}.
\end{equation}
The set of coefficients $\ve{y}$ is computed at once by solving:
\begin{equation}
  \label{eq:715}
  \ve{y} = \arg \underset{\ve{y} \in \Rr^{\text{card}\ca}} {\min}
    \Esp{\prt{\cm(\Ve{X}) - \sum_{\ua \in
          \ca}  y_{\ua} \, \Psi_{\ua}(\Ve{X})}^2}.
\end{equation}
In practice the discretized version of the problem is obtained by
replacing the expectation operator in Eq.(\ref{eq:715}) by the empirical
mean over a sample set:
\begin{equation}
  \label{eq:716}
  \hat{\ve{y}} = \arg \underset{\ve{y} \in \Rr^{\text{card}\ca}} {\min}
  \,     \frac{1}{N} \sum_{i=1}^N \prt{\cm(\Ve{x}^{(i)}) - \sum_{\ua \in
      \ca}  y_{\ua} \, \Psi_{\ua}(\Ve{x}^{(i)})}^2.
\end{equation}
In this expression, $\cx = \acc{\ve{x}^{(i)}, \, i=1\enu n}$ is a sample
set of points called {\em experimental design} (ED) that is typically
obtained by Monte Carlo simulation of the input random vector $\Ve{X} $.
To solve the least-square minimization problem in Eq.(\ref{eq:716}) the
computational model $\cm$ is first run for each point in the ED, and the
results are stored in a vector $ \cy=\acc{y^{(1)} = \cm(\ve{x}^{(1)})
\enu y^{(n)} = \cm(\ve{x}^{(n)})}\tr.$ Then the so-called \textit{information 
matrix} is calculated from the evaluation of the basis polynomials onto each 
point in the ED:
\begin{equation}
  \label{eq:718}
  \imx = \acc{\imx_{ij} \eqdef \Psi_j(\ve{x}^{(i)})\, , \;
    i=1 \enu n, \quad j=1\enu \text{card}~\ca}.
\end{equation}
The solution of the least-square minimization problem finally reads:
\begin{equation}
  \label{eq:719}
  \hat{\ve{y}} = \prt{\imx \tr \imx} ^{-1} \imx\tr \, \cy.
\end{equation}
The points used in the experimental design may be obtained from crude
Monte Carlo simulation. However other types of designs are of common
use, especially Latin Hypercube sampling (LHS), see \citet{McKay1979},
or quasi-random sequences such as the Sobol' or Halton sequence
\citep{Niederreiter1992}. The size of the experimental design is of
crucial importance: it must be larger than the number of unknowns
$\text{card}\ca$ for the problem to be well-posed. In practice we use
the thumb rule $n \approx 2$ - $3\,\text{card}~\ca$ \citep{BlatmanThesis}.

The simple least-square approach summarized above does not allow one to
cope with the curse of dimensionality. Indeed the standard truncation scheme 
requires approximately $ 3 \cdot \binom{d+p}{p}$ runs of the 
original model $\cm(\Ve{x})$, which is in the order of $10^4$ when \eg $d\ge 
15, \,p\ge 5$. However, in practice most of the
problems lead eventually to {\em sparse expansions}, \ie PCE in which
most of the coefficients are zero or negligible. In order to find
directly the significant polynomials and associated coefficients, sparse
PCE have been introduced recently by \citet{BlatmanCras2008,
  BlatmanPEM2010, Bieri:Schwab:2008}. The recent developments make use
of specific selection algorithms which, by solving a penalized
least-square problem, lead by construction to sparse expansions. Of
interest in this chapter is the use of the {\em least-angle regression}
algorithm (LAR, \citet{Efron2004}), which was introduced in the field of
uncertainty quantification by \citet{SudretJCP2011}. Details can be
found in \citet{SudretBookPhoon2015}. Note that other techniques based
on compressive sensing have also been developed recently, see \eg
\citet{Doostan2011,Sargsyan2014,Jakeman2015}.

\subsection{Error estimation}
The truncation of the polynomial chaos expansion introduces an
approximation error which may be computed a posteriori. Based on the
data contained in the experimental design, the {\em empirical error} may
be computed from Eq.(\ref{eq:716}) once least-square minimization
problem has been solved:
\begin{equation}
  \label{eq:720}
  {\vare}_{emp} = \frac{1}{N} \sum_{i=1}^N \prt{\cm(\Ve{x}^{(i)}) - \sum_{\ua \in
      \ca}  \hat{y}_{\ua} \, \Psi_{\ua}(\Ve{x}^{(i)})}^2.
\end{equation}
However, this estimator usually underestimates severely the mean square
error in Eq.(\ref{eq:714}). In particular, if the size $N$ of the
experimental design is close to the number of unknown coefficients
card~$\ca$, the empirical error tends to zero whereas the true mean
square error does not.

A more robust error estimator can be derived based on the {\em
  cross-validation} technique. The experimental design is split into a
training set and a validation set: the coefficients of the expansion are
computed using the training set (Eq.(\ref{eq:716})) whereas the error is
estimated using the validation set. The {\em leave-one-out}
cross-validation is a particular case in which all points but one are
used to compute the coefficients. Setting aside $\ve{x} ^{(i)} \in \cx$,
a PCE denoted by $\mpci(\Ve{X}) $ is built up using the experimental
design $\cx \backslash \ve{x} ^{(i)} \eqdef \acc{\ve{x} ^{(1)} \enu
  \ve{x} ^{(i-1)} , \, \ve{x} ^{(i+1)} \enu \ve{x} ^{(n)}} $ . Then the
error is computed at point $\ve{x}^{(i)}$:
\begin{equation}
  \label{eq:721}
  \Delta_i \eqdef \cm(\ve{x}^{(i)}) -  \mpci(\ve{x}^{(i)}).
\end{equation}
The LOO error is defined by:
\begin{equation}
  \label{eq:722}
  {\vare}_{LOO} = \frac{1}{n}\sum_{i=1}^n \Delta_i^2 =
  \frac{1}{n} \sum_{i=1}^n  \prt{\cm(\ve{x}^{(i)}) -
      \mpci(\ve{x}^{(i)})}^2.
\end{equation}
After some algebra this reduces to:
\begin{equation}
  \label{eq:723}
  \vare_{LOO}=\frac{1}{n}\sum_{i=1}^n  \prt{\frac{\cm(\ve{x}^{(i)}) -
      \mpc(\ve{x}^{(i)}) }{1-h_i}}^2,
\end{equation}
where $h_i$ is the $i$-th diagonal term of matrix $\imx
(\imx\tr\imx)^{-1} \imx\tr$ (matrix $\imx$ is defined in
Eq.(\ref{eq:718})) and $\mpc(\cdot)$ is now the PC expansion built up
from the {\em full} experimental design $\cx$.

As a conclusion, when using a least-square minimization technique to
compute the coefficients of a PC expansion, an {\em a posteriori}
estimator of the mean-square error is readily available. This allows one
to compare PCEs obtained from different truncation schemes and select
the best one according to the leave-one-out error estimate.

\subsection{Post-processing for sensitivity analysis}
\subsubsection{Statistical moments}

The truncated PC expansion $\hat Y = G^{PC}(\Ve{X}) = \sum_{\ua \in \ca} 
\hat{y}_{\ua} \,
\Psi_{\ua}(\Ve{X})$ contains all the information about the statistical
properties of the random output $Y=\cm(\Ve{X} )$. Due to the
orthogonality of the PC basis, mean and standard deviation of $\hat{Y}$ may be
computed directly from the coefficients $\hat{\Ve{y}}$. Indeed, since 
$\Psi_{\ve{0}} \equiv 1$, we
get $\Esp{\Psi_{\ua}(\Ve{X} )} = 0 \quad \forall \, \ua \ne \ve{0}$.
Thus the mean value of $\hat Y$ is the first term of the series:
\begin{equation}
  \label{eq:741}
  \Esp{\hat Y} = \Esp{\sum_{\ua \in \ca} \hat{y}_{\ua} \,
    \Psi_{\ua}(\Ve{X})} ={\hat y}_{\ve{0}}.
\end{equation}
Similarly, due to Eq.(\ref{eq:707}) the variance of $\hat Y$ may be cast
as: 
\begin{equation}
  \label{eq:742}
  \sigma^2_{\hat Y} \eqdef   \Var{\hat Y} = \Esp{\prt{\hat Y - {\hat y}_{\ve{0}}}^2} = \sum_{
    \substack{\ua
      \in \ca \\\ua \ne \ve{0}}} {\hat{y}_{\ua}}^2.
\end{equation}
In other words the mean and variance of the random response may be
obtained by a mere combination of the PCE coefficients once the latter
have been computed.

\subsubsection{Sobol' decomposition and indices}
As already discussed in Chapter~4, global sensitivity analysis is
based on Sobol' decomposition of the computational model $\cm$ (a.k.a
{\em generalized ANOVA decomposition}), which reads \citep{sobol1993}:
\begin{equation}
  \label{eq:743}
  \cm(\ve{x}) = \cm_0 + \sum_{i=1}^d \cm_i(x_i) +
  \sum_{1\leq i < j \leq d} \cm_{ij}(x_i,\,x_j) + \dots +\cm_{12\dots d}(\ve{x}),
\end{equation}
that is, as a sum of a constant $\cm_0$, univariate functions
$\acc{\cm_i(x_i)\,,\, 1\le i\le d}$, bivariate functions\\
$\acc{\cm_{ij}(x_i,x_j)\,,\, 1\le i <j \le d}$, etc.  A recursive
construction is obtained by the following recurrence relationship:
\begin{equation}
  \label{eq:747}
  \begin{split}
    \cm_0 &= \Esp{\cm(\Ve{X} ) } \\
    \cm_i(x_i) &= \Esp{\cm(\Ve{X}) |X_i =x_i} - \cm_0 \\
    \cm_{ij}(x_i,x_j) &= \Esp{\cm(\Ve{X}) |X_i,X_j =x_i,x_j} -
    \cm_i(x_i) -\cm_j(x_j) -\cm_0.
  \end{split}
\end{equation}
Using the {\em set notation} for indices
\begin{equation}
  \label{eq:748}
  A \eqdef \acc{i_1 \enu i_s} \subset \acc{1 \enu d},
\end{equation}
the Sobol' decomposition in Eq.(\ref{eq:743}) reads:
\begin{equation}
  \label{eq:749}
  {\cm}(\ve{x}) = {\cm}_0 + \sum_{\substack{A \subset \acc{1\enu
        d}\\  A \ne \emptyset}} \cm_{A} (\ve{x}_{A}),
\end{equation}
where $\ve{x}_{A}$ is a subvector of $\ve{x} $ which only contains the
components that belong to the index set $A$.  It can be proven that
the summands are orthogonal with each other:
\begin{equation}
  \label{eq:750}
  \Esp{\cm_{A} (\ve{x}_{A }) \, \cm_{B} (\ve{x}_{B })} = 0 \quad
  \forall \; A, B \subset \acc{1\enu d}, \quad A \ne B.
\end{equation} 
Using this orthogonality property, one can 
decompose the variance of the model output
\begin{equation}
  \label{eq:751}
  V \eqdef \Var{Y} =  \Var{\sum_{\substack{A \subset \acc{1\enu
          d}\\  A \ne \emptyset}} \cm_{A} (\ve{x}_A)}
  = \sum_{\substack{A \subset \acc{1\enu
        d}\\  A \ne \emptyset}} \Var{\cm_{A}(\Ve{X}_A)} 
\end{equation}
as the sum of so-called {\em partial variances} defined by:
\begin{equation}
  \label{eq:752}
  V_A \eqdef \Var{\cm_{A}(\Ve{X}_A)} = \Esp{\cm^2_{A}(\Ve{X}_A)}.  
\end{equation}
The Sobol' index attached to each subset of variables $ A \eqdef
\acc{i_1 \enu i_s} \subset \acc{1 \enu d}$ is finally defined by:
\begin{equation}
  \label{eq:753}
  S_A = \frac{V_A}{V} = \frac{\Var{\cm_{A}(\Ve{X}_A)}}{\Var{Y}}.
\end{equation}
{\em First-order} Sobol' indices quantify the portion of the total
variance $V$ that can be apportioned to the sole input variable $X_i$:
\begin{equation}
  \label{eq:754}
  S_i = \frac{V_i}{V} = \frac{\Var{\cm_i(X_i)}}{\Var{Y}}.
\end{equation} 
{\em Second-order} indices quantify the joint effect of variables
$(X_i,X_j)$ that cannot be explained by each single variable separately:
\begin{equation}
  \label{eq:755}
  S_{ij} = \frac{V_{ij}}{V} = \frac{\Var{\cm_{ij}(X_i,X_j)}}{\Var{Y}}.
\end{equation}
Finally, {\em total} Sobol' indices $S^{\mbox{\scriptsize{tot}}}_i$ quantify the total impact of
a given parameter $X_i$ including all of its interactions with other variables. 
They may be computed by the sum of the Sobol' indices of any order that contain 
$X_i$:
\begin{equation}
  \label{eq:756}
  S^{\mbox{\scriptsize{tot}}}_i = \sum_{A \ni i}S_A.
\end{equation}
Amongst other methods, Monte Carlo estimators of the 
various indices are available in the literature and 
thoroughly discussed in (see Variance-based 
sensitivity analysis: Theory and estimation algorithms). 
Their computation usually requires $10^{3-4}$ runs of the 
model $\cm$ {\em for each index}, which leads to a global 
computational cost that is not affordable when $\cm$ is 
expensive-to-evaluate.

\subsubsection{Sobol' indices and PC expansions}
As can be seen by comparing Eqs.(\ref{eq:713}) and (\ref{eq:749}), both
polynomial chaos expansions and Sobol' decomposition are sums of
orthogonal functions. Taking advantage of this property, it is possible
to derive {\em analytic expressions} for Sobol' indices based on a PC
expansion, as originally shown in \citet{SudretCSM2006,SudretRESS2008b}.
For this purpose let us consider the set of multivariate polynomials
$\Psi_{\ua}$ which depend {\em only} on a subset of variables $A =
\acc{i_1 \enu i_s} \subset \acc{1 \enu d}$:
\begin{equation}
  \label{eq:757}
  \ca_A = \acc{\ua \in \ca \,:\;   \alpha_k \ne
    0 \text{~~if and only if~~}k\in A}.
\end{equation}
The union of all these sets is by construction equal to $\ca$. Thus we
can reorder the terms of the truncated PC expansion so as to exhibit the
Sobol' decomposition:
\begin{equation}
  \label{eq:758}
  \mpc(\ve{x}) = y_0 + \sum_{\substack{A \subset \acc{1\enu
        d}\\  A \ne \emptyset}} \mpc_{A} (\ve{x}_{A})
  \qquad \text{where} \quad
  \mpc_{A} (\ve{x}_{A}) \eqdef \sum_{\ua \in
    \ca_{A}} y_{\ua} \, \Psi_{\ua}(\Ve{x}).
\end{equation}
Consequently, due to the orthogonality of the PC basis, the partial
variance $V_A$ reduces to:
\begin{equation}
  \label{eq:759}
  V_A = \Var{\mpc_{A} (\ve{X}_{A}) } = \sum_{\ua \in
  \ca_{A}} y^2_{\ua} .
\end{equation}
In other words, from a given PC expansion, the Sobol' indices {\em at
  any order} may be obtained by a mere combination of the squares of the
coefficients. More specifically, the PC-based estimator of the
first-order Sobol' indices read:
\begin{equation}
  \label{eq:760}
  \hat S_i = \frac{\dsp{\sum_{\ua \in \ca_i} \hat
      y_{\ua}^2}}{\dsp{\sum_{\ua   \in \ca \;,\; \ua \ne \ve{0}} \hat y_{\ua}^2}} \qquad
  \text{where}  \quad \ca_i = \acc{
    \ua \in \ca \,:\; \alpha _i >  0 \,,\, \alpha_{j \ne i}= 0 } .
\end{equation}
and the total PC-based Sobol' indices read:
\begin{equation}
  \label{eq:761}
  \hat  S^{\mbox{\scriptsize{tot}}}_i = \frac{\dsp{\sum_{\ua \in \ca^{\mbox{\scriptsize{tot}}}_i} \hat
      y_{\ua}^2}}{\dsp{\sum_{\ua   \in \ca \;,\; \ua \ne \ve{0}} \hat
      y_{\ua}^2}}   \qquad \ca^{\mbox{\scriptsize{tot}}}_i = \acc{ \ua
    \in \ca \,:\; \alpha _i > 0 }.
\end{equation}

\subsection{Summary}
Polynomial chaos expansions allow one to cast the random response
$\cm(\Ve{X} ) $ as a truncated series expansion. By selecting an
orthonormal basis w.r.t. the input parameter distributions, the
corresponding coefficients can be given a straightforward
interpretation: the first coefficient $y_0$ is the mean value of the
model output whereas the variance is the sum of the squares of the remaining
coefficients. Similarly, the Sobol' indices are obtained by summing up
the squares of suitable coefficients. Note that in low dimension
($d<10$) the coefficients can be computed by solving a mere ordinary
least-square problem. In higher dimensions advanced techniques leading
to sparse expansions must be used to keep the total computational cost
(measured in terms of the size $N$ of the experimental design)
affordable. Yet the post-processing to get the Sobol' indices from the
PCE coefficients is independent of the technique used.

\section{Gaussian process-based sensitivity analysis}
\label{sec:7-03}

\subsection{A short introduction to Gaussian processes}

Let us consider a probability space $(\Omega_Z, 
\mathcal{F}_Z, \mathbb{P}_Z)$, a measurable space 
$(\mathcal{S}, \mathcal{B}(\mathcal(S))$ and an arbitrary 
set $T$.  A stochastic process $Z(\ve{x})$, $\ve{x} \in 
T$,  is Gaussian if and only if for any finite subset $C 
\subset T$, the collection  of random variables $Z(C)$  has 
a Gaussian joint distribution. 
In our framework, $T$  and $S$ represent the input and the output spaces. Therefore, we have $T = \mathbb{R}^d$ and $S = \mathbb{R}$.

\begin{figure}[!ht]
  \begin{center}
    \includegraphics[width=0.49\linewidth,angle=0]{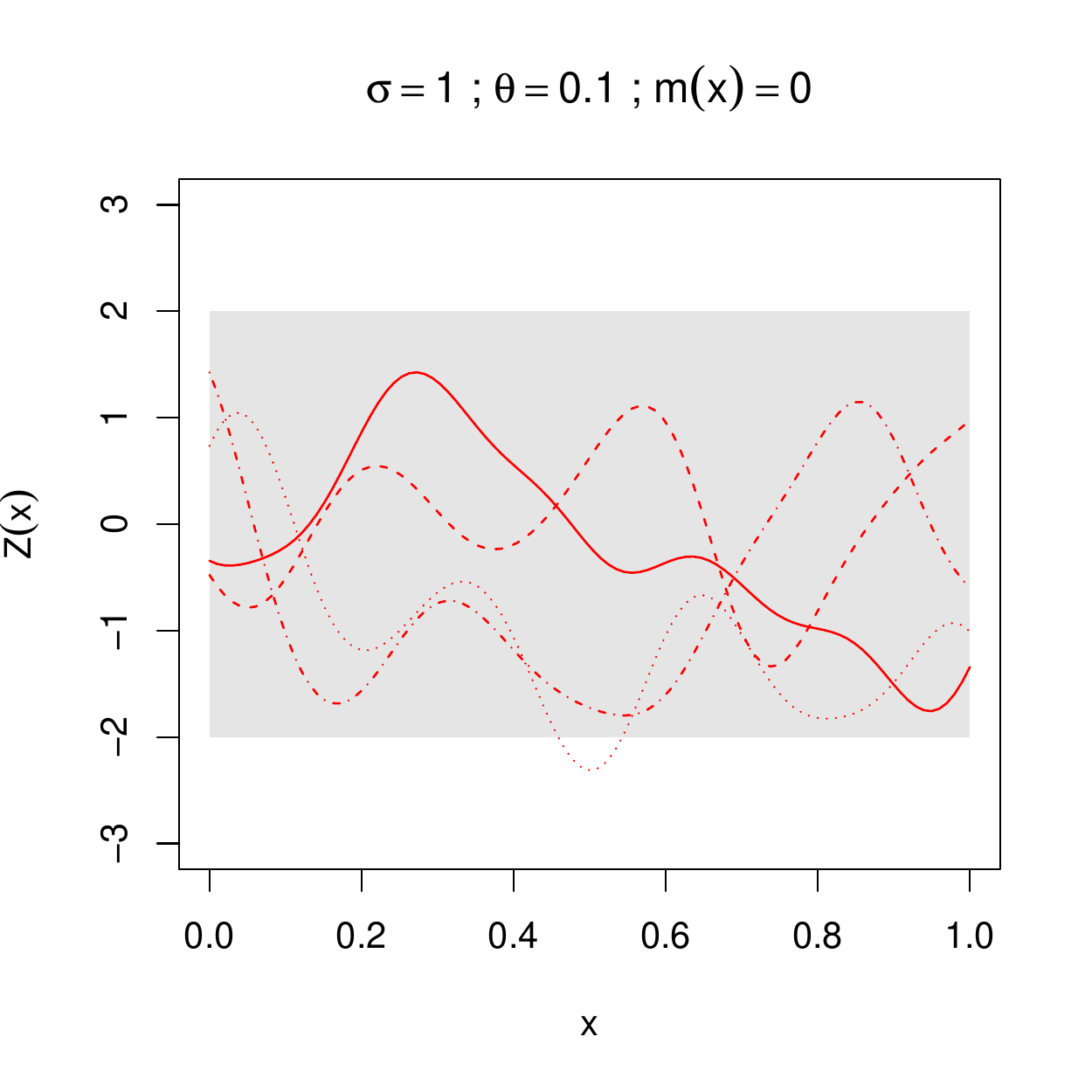}
    \includegraphics[width=0.49\linewidth,angle=0]{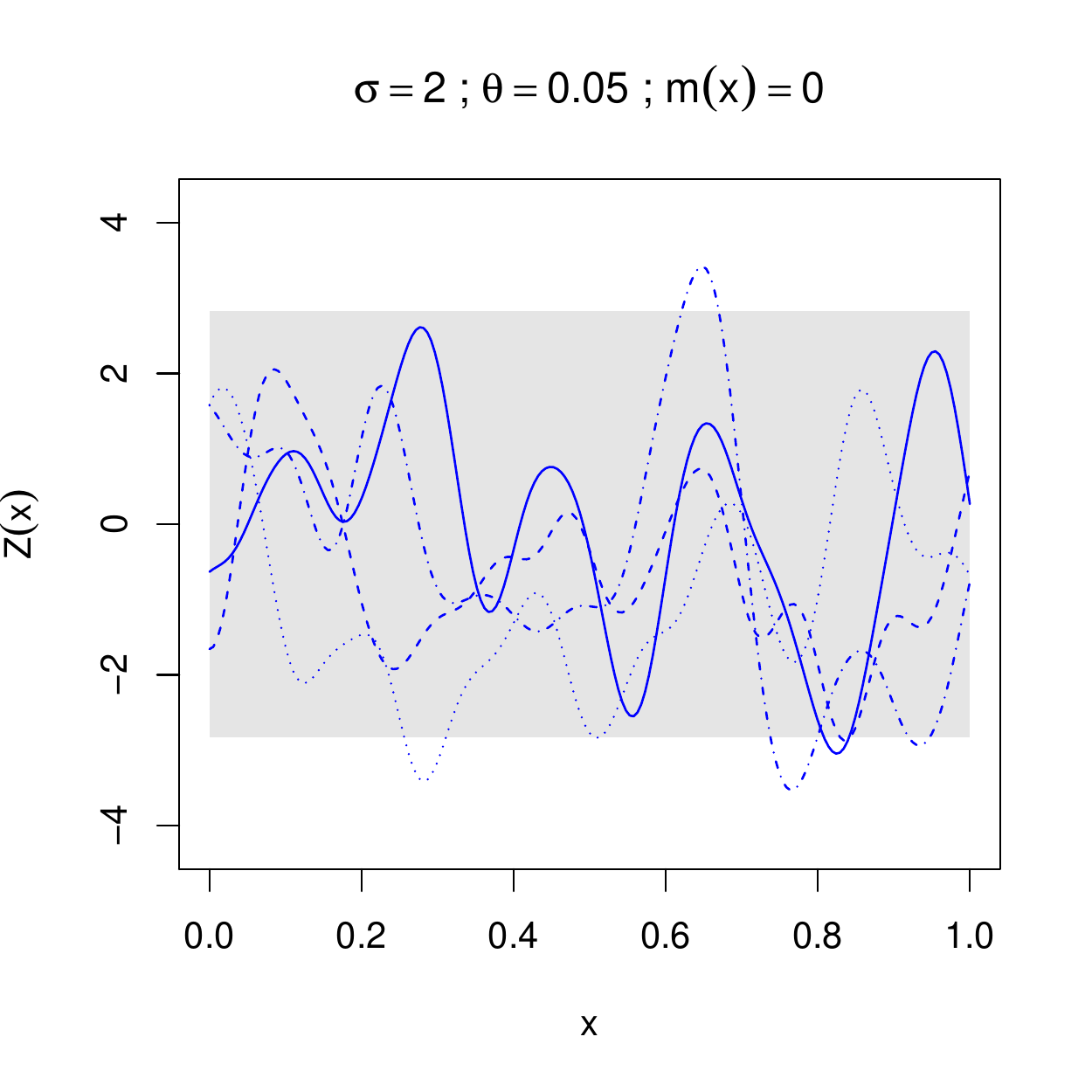}
        \includegraphics[width=0.49\linewidth,angle=0]{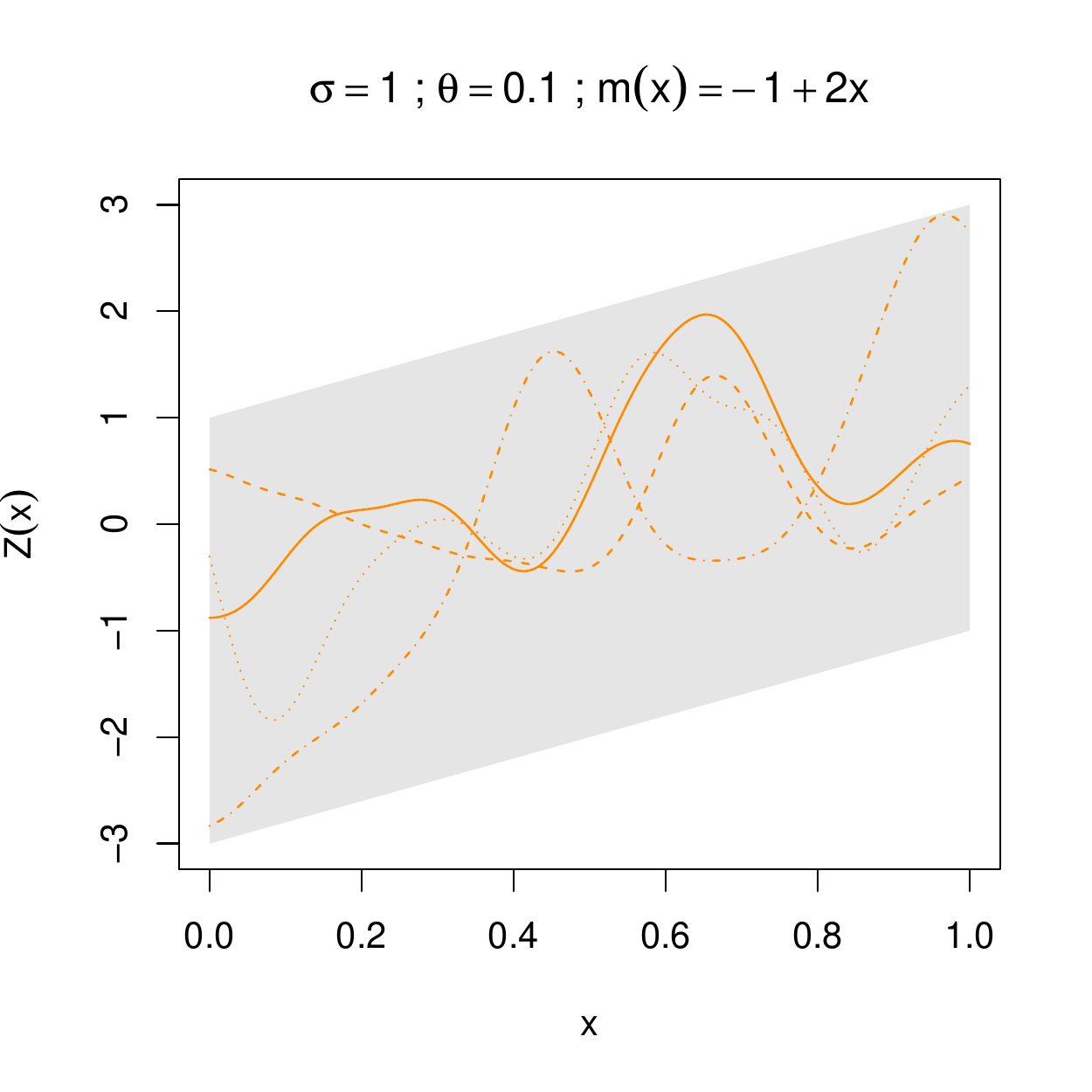}
    \caption{Examples of Gaussian process realizations 
    with   squared exponential kernels and different means. 
    The shaded areas represent  the point-wise 95\% 
    confidence intervals.}
    \label{Zexample}
  \end{center}
\end{figure}

A Gaussian process is entirely specified by its mean $m(\ve{x}) = \mathbb{E}_Z[Z(\ve{x})]$  and covariance function $k(\ve{x},\ve{x}') =   \mathrm{cov}_Z(Z(\ve{x}),Z(\ve{x}'))$ 
where $\mathbb{E}_Z$ and $\mathrm{cov}_Z$ denote  the expectation and the covariance  with respect to $(\Omega_Z, \mathcal{F}_Z, \mathbb{P}_Z)$.
The covariance function $k(\ve{x},\ve{x}')$ is a positive 
definite kernel. It is often considered stationary \ie{} 
$k(\ve{x},\ve{x}')$ is a function of $\ve{x}-\ve{x}'$.
The covariance kernel is the most important term of a Gaussian process regression. Indeed, it controls the smoothness and the scale of the approximation. A popular choice for $k(\ve{x},\ve{x}')$ is the stationary isotropic squared exponential kernel defined as :
\begin{displaymath}
k(\ve{x},\ve{x}') = \sigma^2 \mathrm{exp} \left( -\frac{1}{2 \ttheta^2} ||\ve{x} - \ve{x}'||^2 \right).
\end{displaymath}
It is parametrized by the parameter $\ttheta$ -- also 
called characteristic length scale or correlation length -- 
and the variance parameter $\sigma^ 2$. We give in Figure 
\ref{Zexample} examples of realizations of Gaussian 
processes with  stationary isotropic squared exponential 
kernels.

 We observe that $m(\ve{x})$ is the trend around which the realizations vary,  $\sigma^ 2$ controls the range of their variation  and  $\ttheta$ controls their oscillation frequencies. We highligh that Gaussian processes with squared exponential covariance kernels are infinitely differentiable almost surely. As mentioned in  \citep{S99}, this  choice of kernel can be unrealistic due to its strong regularity.

\subsection{Gaussian process regression models}
\label{sec:7-03.1}
The principle of Gaussian process regression is to consider that the
prior knowledge about the computational model $G(\ve{x})$, $\ve{x} \in \R^d$, 
can be \textbf{modeled} by a Gaussian process $Z(\ve{x})$ with a 
mean denoted by 
$m(\ve{x})$ and a
covariance kernel denoted by $k(\ve{x},\ve{x}')$.  
Roughly speaking, we consider that the true response is a realization of $Z(\ve{x})$.
Usually, the mean and the covariance are 
parametrized
as follows:
\begin{equation}
  m(\ve{x}) = \ff\tr(\ve{x}) \bbeta,
\end{equation}
and
\begin{equation}
k(\ve{x},\ve{x}') = \sigma^2 r(\ve{x},\ve{x}';\ttheta),
\end{equation}
where $\ff\tr(\ve{x})$ is a vector of $p$ prescribed functions and
$\bbeta$, $\sigma^2$ and $\ttheta$ have to be estimated.  The mean
function $m(\ve{x})$ describes the trend and the covariance kernel
$k(\ve{x},\ve{x}')$ describes the regularity and characteristic length
scale of the model.

\subsubsection{Predictive distribution} 

Consider an experimental design
$\cx = \acc{\ve{x}^{(1)}, \dots, \ve{x}^{(n)}}$, $\ve{x}^{(i)} \in\R^d$, and
the corresponding model responses $\cy = G(\cx)$. The predictive
distribution of $G(\ve{x})$ is given by:
\begin{equation}
  \label{book4Chapter7_GP_1}
  [Z(\ve{x})|Z(\cx) = \cy,  \sigma^2, \ttheta] \sim \GP{m_n(\ve{x})}{k_n(\ve{x},\ve{x}')},
\end{equation} 
where
\begin{equation}
  \label{book4Chapter7_GP_3}
  m_n(\ve{x}) = \ff\tr(\ve{x}) \bar{\bbeta} + \rr\tr(\ve{x})\K^{-1} \left( \cy - \FF \bar 
    \bbeta\right),
\end{equation}
\begin{equation}
  \label{book4Chapter7_GP_2}
  k_n(\ve{x},\ve{x}') = \sigma^2 \left( 1 - \begin{pmatrix} \ff\tr(\ve{x}) & \rr\tr(\ve{x}) 
    \end{pmatrix}
    \begin{pmatrix} 0 & \FF\tr \\ \FF & \K \end{pmatrix} \begin{pmatrix}
      \ff(\ve{x}') \\ \rr(\ve{x}')  
    \end{pmatrix}\right),
\end{equation}
\begin{figure}[!ht]
  \begin{center}
   \includegraphics[width=0.65\linewidth,angle=0, trim = 0 0
  0 50, clip]{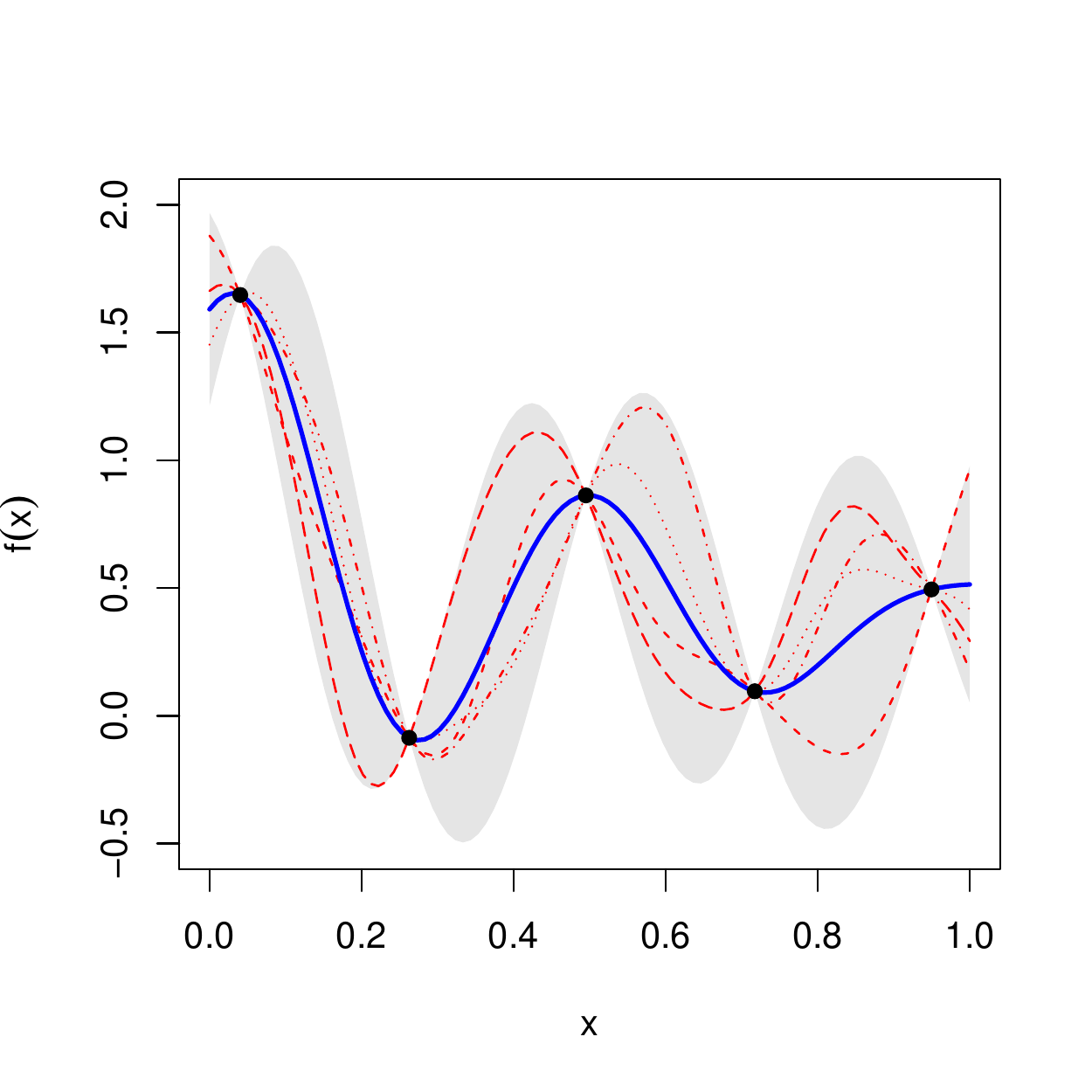}
    \caption{Examples of predictive distribution. The solid 
    line represents the mean of the predictive 
    distribution, the non-solid lines represent some of 
    its   realizations and the shaded area represents the 
    95\% confidence intervals based on the variance of 
    the predictive distribution.}
    \label{exemple_cond}
  \end{center}
\end{figure}
In these expressions $\K = [r(\ve{x}^i,\ve{x}^j;
\ttheta)]_{i,j=1,\dots,n}$, $\rr(\ve{x}) =
[r(\ve{x},\ve{x}^{(i)};\ttheta)]_{i=1,\dots,n}$, $\FF =
[\ff\tr(\ve{x}^{(i)})]_{i=1,\dots,n}$ and
\begin{equation}
\label{book4Chapter7_GP_2bis}
  \bar{\bbeta} = \left(\FF\tr \K^{-1} \FF \right)^{-1} \FF\tr \K^{-1} \cy.
\end{equation}
The term $\bar{\bbeta}$ denotes the posterior distribution mode of
$\bbeta$ obtained from the improper non-informative prior distribution
$\pi(\bbeta) \propto 1$ \citep{robert2007bayesian}. 

\emph{Remark.} The predictive distribution is given by the 
Gaussian process $Z(x)$ conditioned by the known 
observations $ \cy$.  The Gaussian process regression 
metamodel is given by the  conditional expectation 
$m_n(x)$    and its mean squared error is given by the 
conditional variance $ k_n(\ve{x},\ve{x}) $.  An 
illustration  of $m_n(x)$ and $ k_n(\ve{x},\ve{x}) $ is 
given in Figure \ref{exemple_cond}.\\

The reader can note that the predictive distribution (\ref{book4Chapter7_GP_1}) integrates
the posterior distribution of $\bbeta$. However, the hyper-parameters
$\sigma^2$ and $\ttheta$ are not known in practice and shall be
estimated with the maximum likelihood method
\citep{harville1977maximum,S03} or a cross-validation strategy
\citep{bachoc2013cross}. Then, their estimates are plugged in the
predictive distribution. The restricted maximum likelihood estimate of
$\sigma^2$ is given by:
\begin{equation}
\hat{\sigma}^2 = \frac{(\cy - \FF\bar \bbeta)\tr\K^{-1}(\cy 
- \FF\bar \bbeta)}{n-p}.
\end{equation}
Unfortunately, such a closed form expression does not exist for $\ttheta$ and 
it has to be numerically estimated.

\emph{Remark.} Gaussian process regression can easily be 
extended to the case of noisy observations. Let us suppose 
that  $\cy$ is tainted by a white Gaussian noise 
$\varepsilon$ :
\begin{displaymath}
\cy_\mathrm{obs} = \cy + \sigma_\varepsilon (\ve{\cx}) \varepsilon.
\end{displaymath}
The term $\sigma_\varepsilon (\ve{\cx})$ represents the 
standard deviation of the observation noise. The mean and 
the covariance of the predictive distribution $ 
[Z(\ve{x})_\mathrm{obs}|Z(\cx) = \cy_\mathrm{obs},  
\sigma^2, \ttheta] $ is then obtained by replacing in 
Equations (\ref{book4Chapter7_GP_3}), 
(\ref{book4Chapter7_GP_2}) and 
(\ref{book4Chapter7_GP_2bis}) the correlation matrix $\K$ 
by $\sigma^2 \K + \mathbf{\Delta}_\varepsilon$ where  
$\mathbf{\Delta}_\varepsilon$ is a diagonal matrix given by 
:
\begin{displaymath}
\mathbf{\Delta}_\varepsilon =
 \begin{pmatrix} 
\sigma_\varepsilon (\ve{x}^{(1)} ) & & &  \\
 & \sigma_\varepsilon (\ve{x}^{(2)} )  & & \\
 & &  \ddots & \\
 & & & \sigma_\varepsilon (\ve{x}^{(n)} ) \\
 \end{pmatrix}  .
\end{displaymath}
We emphasize that the closed form expression for the restricted maximum likelihood estimate of $\sigma^2$ does not exist anymore. Therefore, this parameter has to be numerically estimated.

\subsubsection{Sequential design} 
To improve the global accuracy of the GP model, it is usual to augment the initial design set $\cx$ 
with new points.
An important feature of Gaussian process regression is that it provides an  estimate of the model 
mean-square error through the term $k_n(\ve{x},\ve{x}')$ (\ref{book4Chapter7_GP_2}) which can be 
used to select these new points. The 
most common but not efficient 
sequential criterion 
consists in adding the point $\ve{x}^{(n+1)}$ where the mean-square error is the largest:
\begin{equation}
\ve{x}^{(n+1)} = \arg \max_{\ve{x}} k_n(\ve{x},\ve{x}).
\end{equation}
More efficient criteria can be found in \citet{Bat96,Klei06,LoicSequential}.

\subsubsection{Model selection} 
To build up a GP model, the user has to
make several choices. Indeed, the vector of functions $\ff(\ve{x})$
and the class of the correlation kernel $r(\ve{x},\ve{x}';\ttheta)$ need
to be set (see \citet{R06} for different examples of correlation
kernels). These choices and the relevance of the model are tested a
posteriori with a validation procedure. If the number $n$ of
observations is large, an external validation may be performed 
on a test set. Otherwise, a cross-validation procedure may be used. An
interesting property of GP models is that a closed form expression
exists for the cross-validation predictive distribution, see for
instance \citet{Dub83}. 
It allows for deriving efficient methods of parameter estimation
\citep{bachoc2013cross} or sequential design \citep{LoicSequential}.\\

Some usual stationary covariance kernel are listed below.

\begin{itemize}
\item[] \textbf{The squared exponential covariance function.} The form of this kernel is given by:
\begin{displaymath}
k(\ve{x},\ve{x}') = \sigma^2 \mathrm{exp} \left( -\frac{1}{2 \ttheta^2} ||\ve{x} - \ve{x}'||^2 \right).
\end{displaymath}
This covariance function corresponds to Gaussian processes which are
infinitely differentiable in mean square and almost surely.  We
illustrate in Figure \ref{ex_gauss_kern} the 1-dimensional squared
exponential kernel with different correlation lengths and examples of
resulting Gaussian process realizations.
\begin{figure}[!ht]
  \begin{center}
    \includegraphics[width=0.45\textwidth, trim = 100 200 100 200,
    clip]{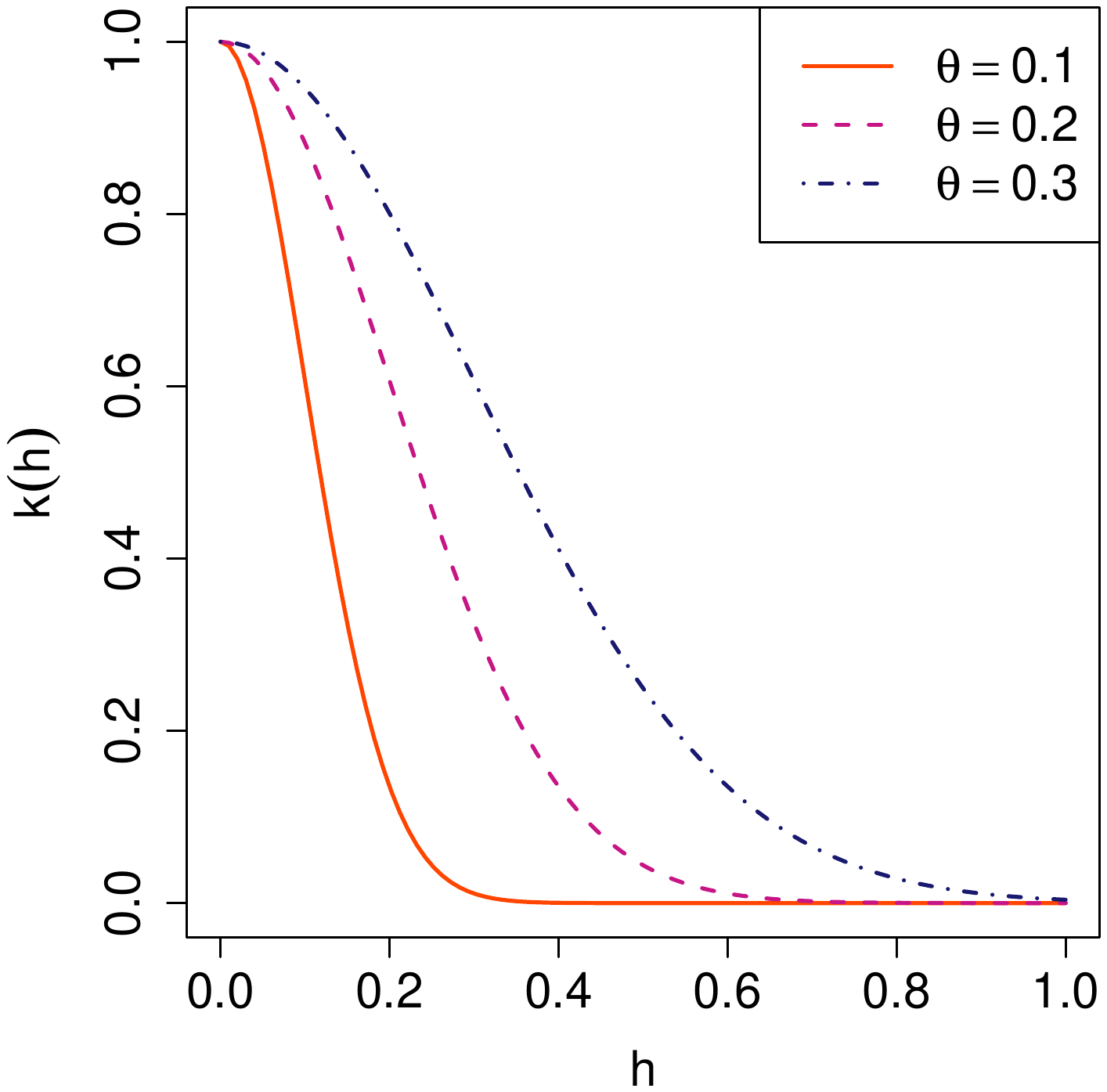}
    \includegraphics[width=0.45\textwidth, trim = 100 200 100 200,
    clip]{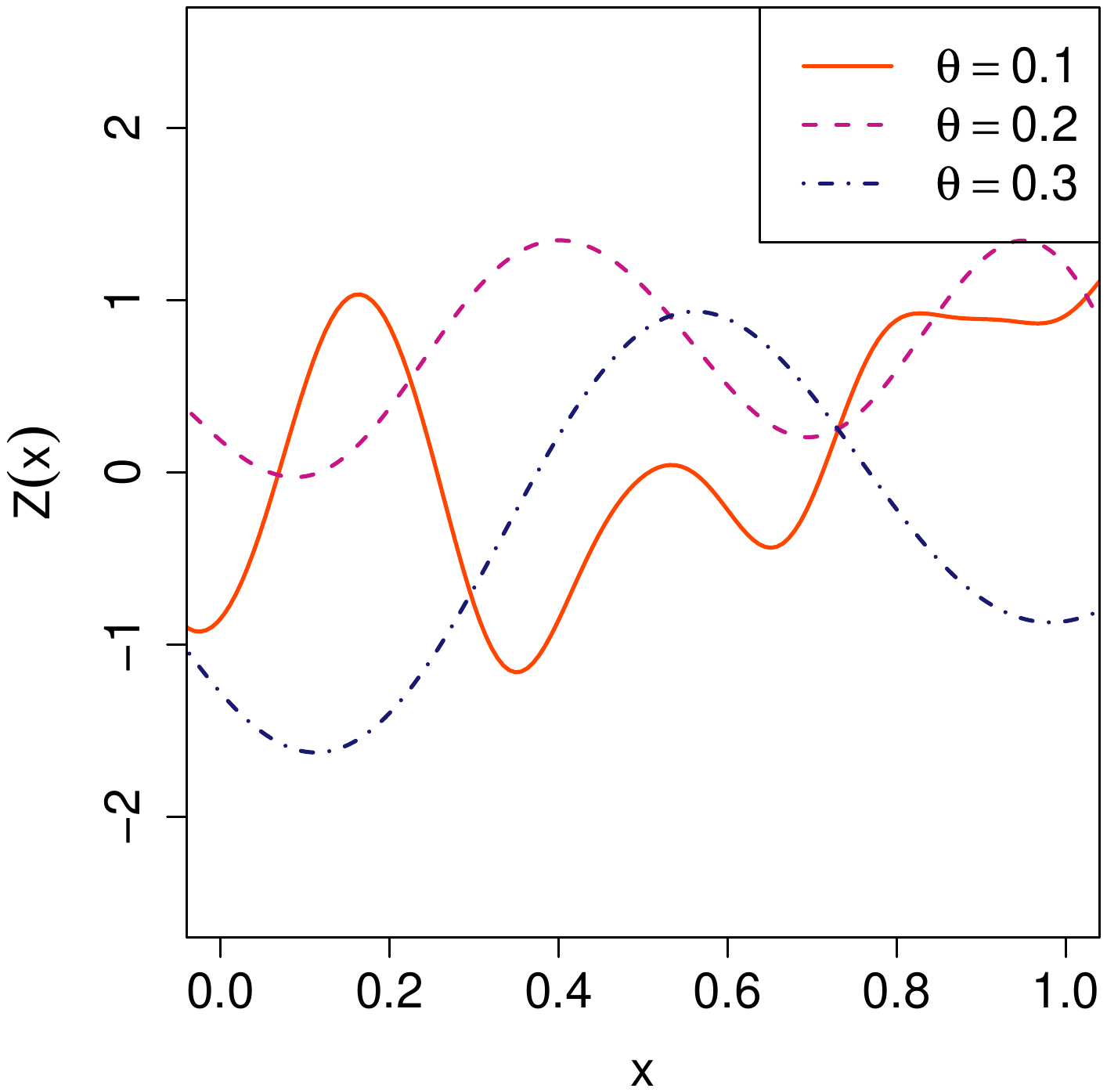}
    \caption{The squared exponential kernel in function of $h = \ve{x} - \ve{x}'$ with different correlation lengths $\ttheta$ and examples of resulting Gaussian process realizations.}
    \label{ex_gauss_kern}
  \end{center}
\end{figure}
\item[] \textbf{The $\nu$-Mat\'ern covariance function.}  This covariance kernel is defined as follow (see \citep{S99}):
\begin{displaymath}
k_\nu(h) = \frac{2^{1- \nu}}{\Gamma(\nu)} \left( \frac{\sqrt{2} ||h||}{\ttheta} \right)^\nu K_\nu \left( \frac{\sqrt{2} \nu ||h||}{\ttheta}\right),
\end{displaymath}
where $\nu$ is the regularity parameter,  $K_\nu$ is a modified Bessel Function and $\Gamma$ is the Euler Gamma function. 
A Gaussian process with a $\nu$-Mat\'ern  covariance kernel is $\nu$-H\"older continuous in mean square and $\nu'$-H\"older continuous almost surely with $\nu' < \nu$.
Three popular choice of  $\nu$-Mat\'ern covariance kernels are the ones for $\nu = 1/2$, $\nu = 3/2$ and $\nu = 5/2$ :
\begin{displaymath}
k_{\nu=1/2}(h) = \exp \left( - \frac{||h||}{\ttheta} \right),
\end{displaymath}
\begin{displaymath}
k_{\nu=3/2}(h) = \left( 1 + \frac{\sqrt{3} ||h||}{\ttheta} \right) \exp \left( - \frac{ \sqrt{3}||h||}{\ttheta} \right),
\end{displaymath}
and
\begin{displaymath}
k_{\nu=5/2}(h) = \left( 1 + \frac{\sqrt{5} ||h||}{\ttheta} +  \frac{ {5} ||h||^2}{3 \ttheta^2} \right) \exp \left( - \frac{ \sqrt{5}||h||}{\ttheta} \right).
\end{displaymath}
We illustrate in Figure \ref{ex_matern_kern} the 1-dimensional $\nu$-Mat\'ern  kernel for different values of $\nu$.
\begin{figure}[h]
  \begin{center}
    \includegraphics[width=0.45\linewidth, trim = 100 200 100 200,
    clip]{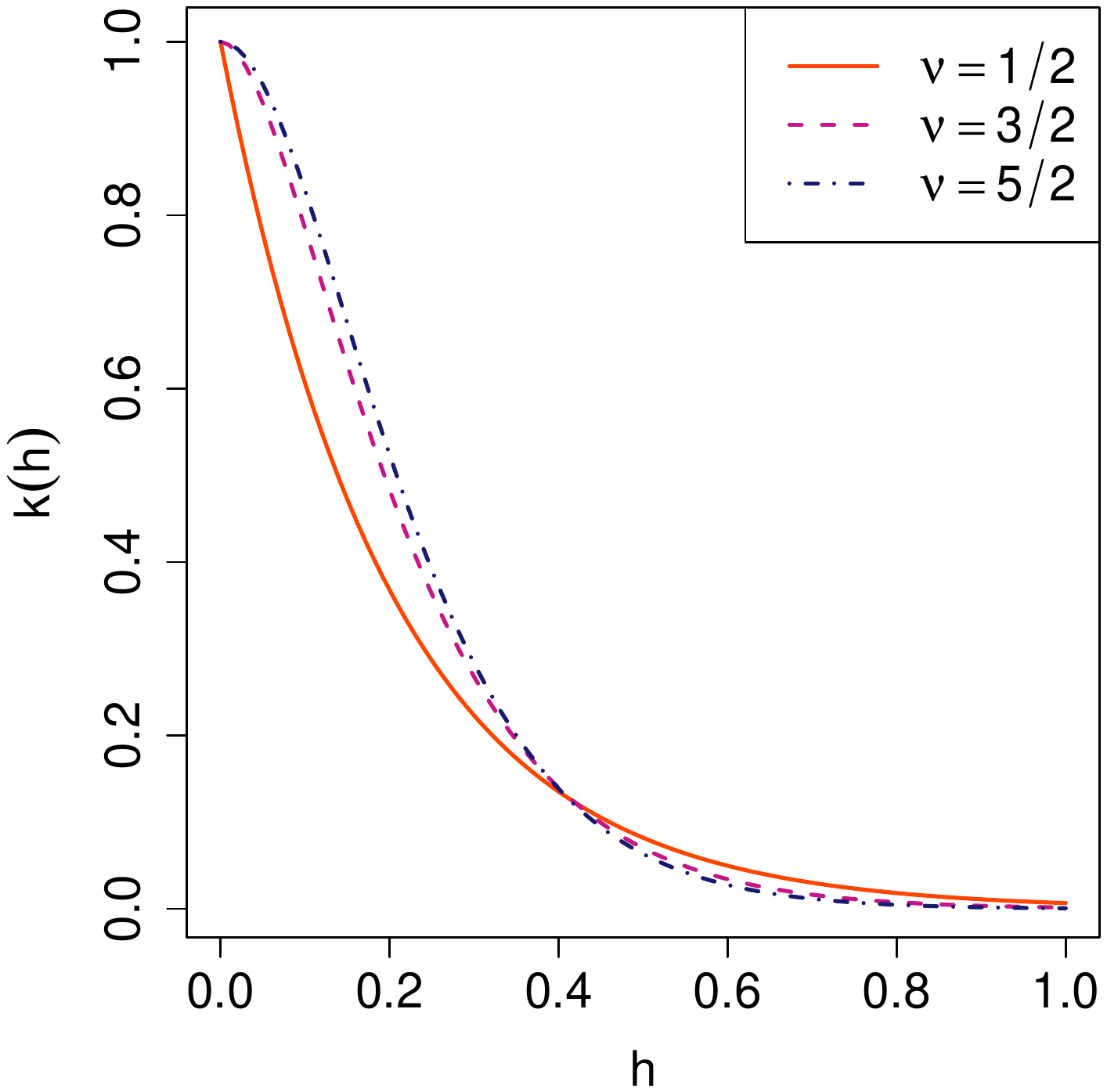}
    \includegraphics[width=0.45\linewidth,trim = 100 200 100 200,
    clip]{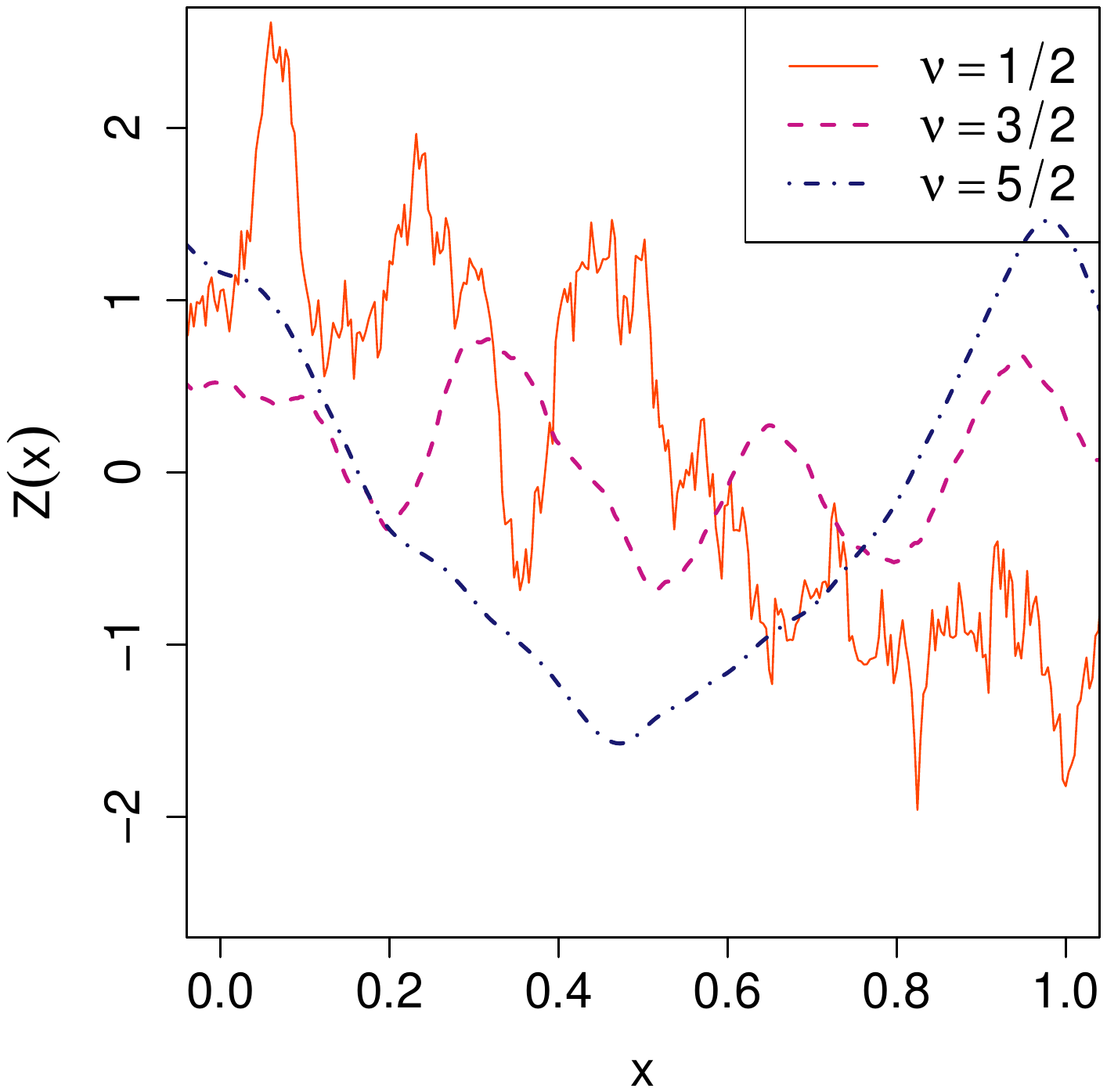}
    \caption{The  $\nu$-Mat\'ern  kernel in function of $h = \ve{x} - \ve{x}'$ with different regularity parameters $\nu$ and examples of resulting Gaussian process realizations.}
    \label{ex_matern_kern}
  \end{center}
\end{figure}
\item[]  \textbf{The $\gamma$-exponential covariance function.} This kernel is defined as follow:
\begin{displaymath}
k_{\gamma}(h) = \exp \left( - \left( \frac{||h||}{\ttheta} \right)^\gamma \right).
\end{displaymath}
For $\gamma < 2$ the corresponding Gaussian process are not differentiable in mean square  whereas for $\gamma = 2$ is is infinitely differentiable (it corresponds to the squared exponential kernel). 
We illustrate in Figure \ref{ex_gexpo_kern} the 
1-dimensional $\gamma$-exponential  kernel for different 
values of $\gamma$.
\begin{figure}[!ht]
  \begin{center}
    \includegraphics[width=0.45\linewidth,trim = 100 200 100 200,
    clip]{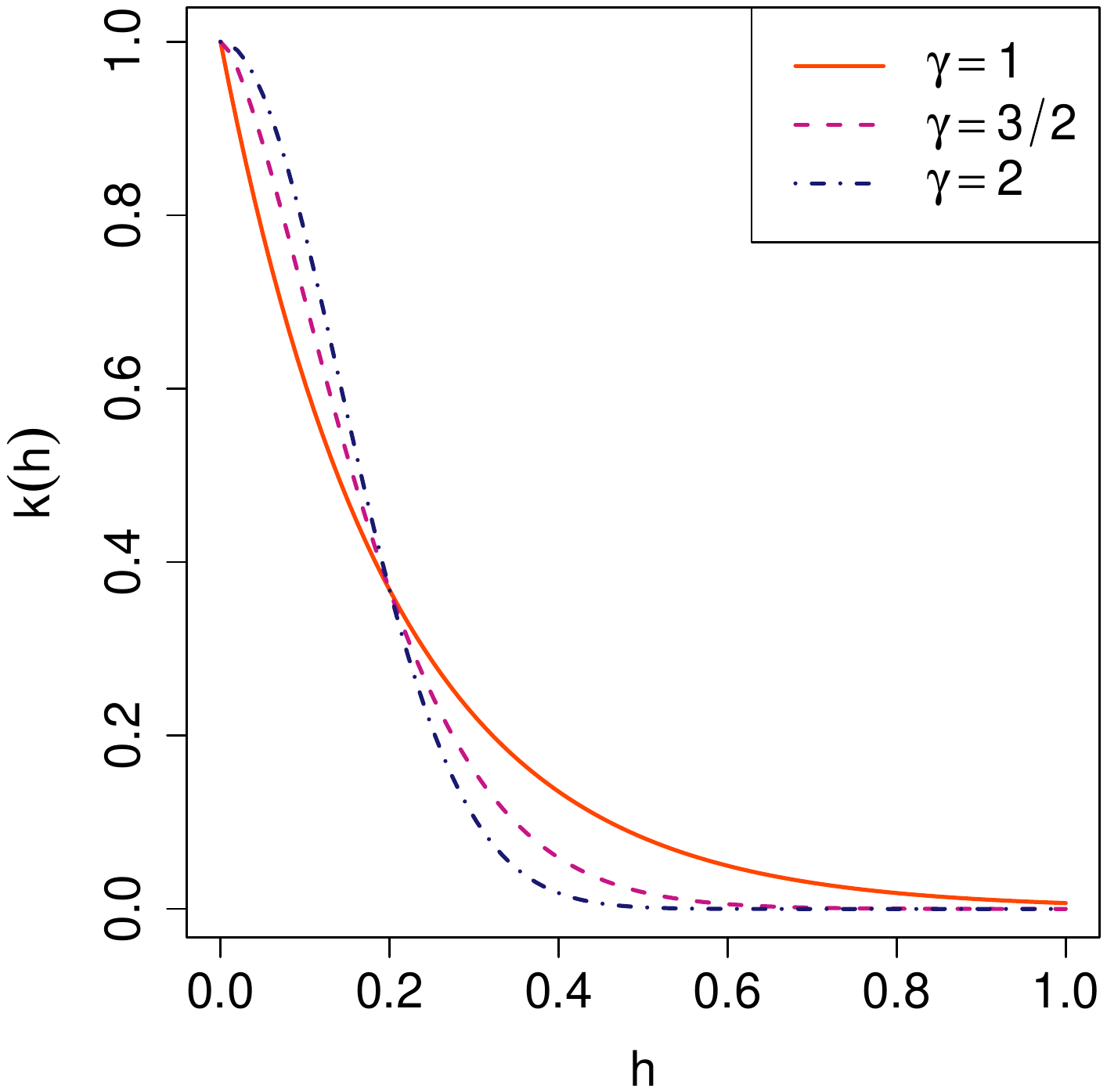}
    \includegraphics[width=0.45\linewidth, trim = 100 200 100 200,
    clip]{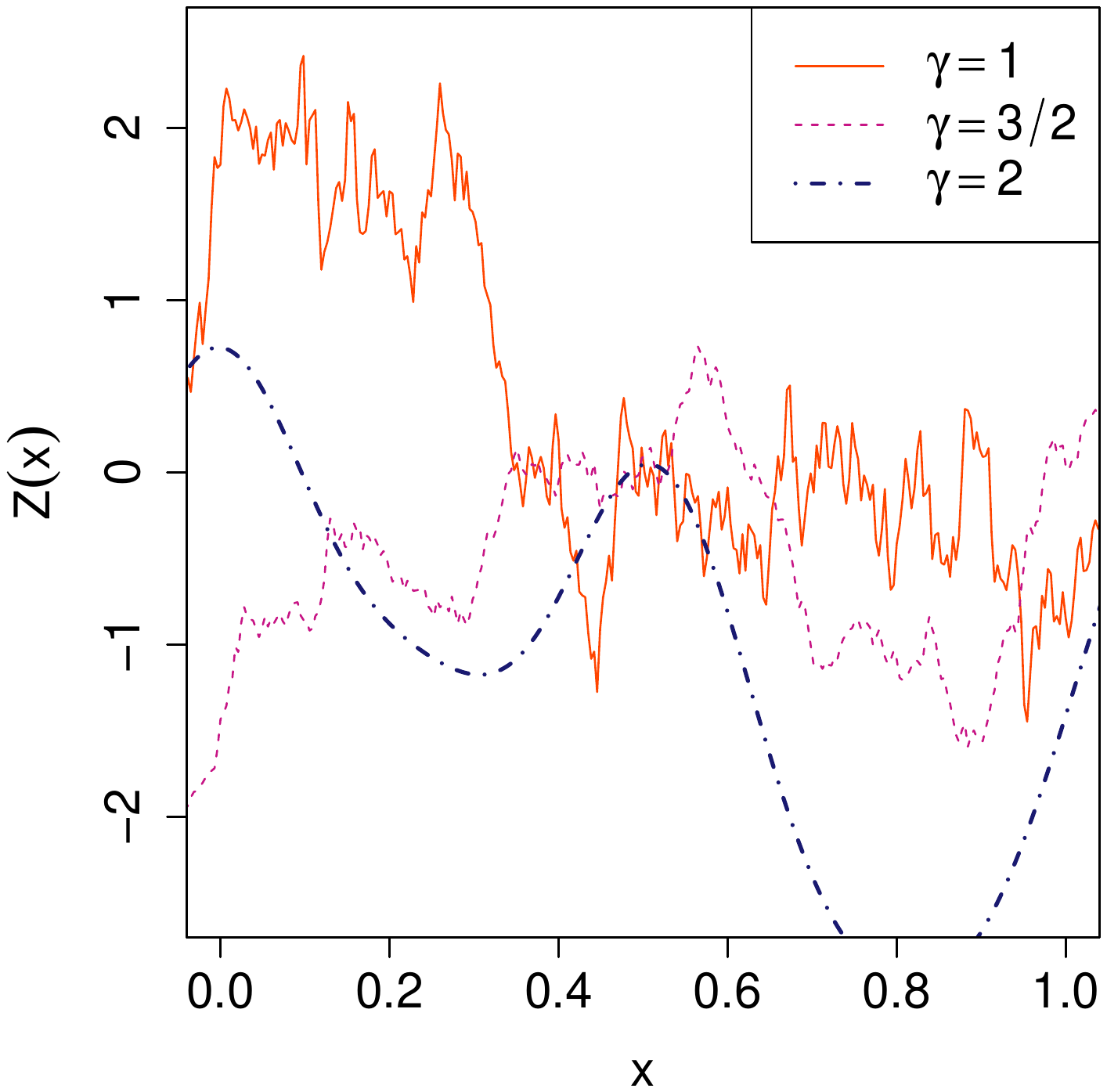}
    \caption{The  $\gamma$-exponetial   kernel in function of $h = \ve{x} - \ve{x}'$ with different regularity parameters $\gamma$  and examples of resulting Gaussian process realizations.}
    \label{ex_gexpo_kern}
  \end{center}
\end{figure}
\end{itemize}

\subsubsection{Sensitivity analysis}
To perform a sensitivity analysis from a GP model, two approaches are 
possible. 
The first one consists in substituting the true model $G(\ve{x})$ with the mean 
of the conditional Gaussian process $m_n(\ve{x})$ in  
(\ref{book4Chapter7_GP_3}). However, it may provide biased sensitivity index 
estimates. Furthermore it does not allow one to quantify 
the error on the  
sensitivity indices due to the metamodel approximation.
The second one consists in substituting $G(\ve{x})$ by a Gaussian process 
$Z_n(\ve{x})$ having the predictive distribution $[Z(\ve{x})|Z(\cx) = \cy,   
\sigma^2, \ttheta]$ shown in (\ref{book4Chapter7_GP_1}). 
This approach makes it possible to quantify the uncertainty due to the 
metamodel approximation and allows for building unbiased index estimates.

\subsection{Main effects visualization}
\label{sec:7-03.2}

From now on, the input parameter $\ve{x} \in \R^d$ is considered as a
random input vector $\X = (X_1,\dots,X_d)$ with independent components.
Before focusing on variance-based sensitivity indices, the inference
about the main effects is studied in this section. Main effects are a powerful
tool to visualize the impact of each input variable on the model output (see 
\eg \citet{Oak04}).  The main effect of the group of input variables $\X_A,
~A\subset \acc{1,\dots,d}$ is defined by $\E{ G(\X) | \X_A}$.  Since the
original model $G$ may be time-consuming to evaluate, it is substituted
for by its approximation, \textit{i.e.} $G(\Ve{X} ) \approx \E{ Z_n(\X)
  | \X_A}$, where $Z_n(\ve{x}) \sim [Z(\ve{x})|Z(\cx) = \cy, \sigma^2,
\ttheta]$.  Since $\E{ Z_n(\X) | \X_A}$ is a linear transformation of
the Gaussian process $Z_n(\ve{x})$, it is also a Gaussian process. The
expectations, variances and covariances with respect to the posterior
distribution of $[Z(\ve{x})|Z(\cx) = \cy, \sigma^2, \ttheta]$ are
denoted by $\EZ{.}$, $\VZ{.}$ and $\CovZ{.}{.}$. Then, we have:
\begin{equation}
\E{ Z_n(\X) | \X_A} \sim \GP{\E{ m_n(\X) | \X_A}}{\E{\E{k_n(\X,\X')|\X_A}|\X'_A}}.
\end{equation}
The term $\E{ m_n(\X)}$ represents the approximation of $\E{ G(\X) |
  \X_A}$ and $\E{\E{k_n(\X,\X')|\X_A}|\X'_A}$ is the mean-square error
due to the metamodel approximation.  Therefore, with this method one
can quantify the error on the main effects due to the metamodel
approximation.  For more detail about this approach, the reader is
referred to \citet{Oak04,Marrel2009}.

\subsection{Variance of the main effects}
\label{sec:7-03.3}

Although the main effect enables one to visualize the impact of a group of 
variables on the 
model output, it does not quantify it. To perform such an analysis, consider the variance of the 
main effect:
\begin{equation}
V_A = \V{\E{ Z_n(\X) | \X_A}},
\end{equation}
or its normalized version which corresponds to the Sobol' index:
\begin{equation}\label{book4Chapter7_GP_5}
S_A = \frac{V_A}{V}= \frac{\V{\E{ Z_n(\X) | \X_A}}}{\V{Z_n(\X)}}.
\end{equation}
Sobol' indices are the most popular measures to carry out a sensitivity
analysis since their value can easily be interpreted as the part of the
total variance due to a group of variables. However, in contrary to the
partial variance $V_A$, it does not provide information about the order of 
magnitude of the contribution to the model output variance of variable group 
$\X_A$.

\subsubsection{Analytic formulae}
The above indices are studied in \citet{Oak04} where the estimation of
$V_A$ and $V$ is performed separately. Indeed, computing the Sobol'
index $S_A$ requires considering the joint distribution of $V_A$ and
$V$, which makes it impossible to derive analytic formulae.  According
to \citet{Oak04}, closed form expressions in terms of integrals can be
obtained for the two quantities $\EZ{V_A}$ and $\VZ{V_A}$.  The quantity
$\EZ{V_A}$ is the sensitivity measure and $\VZ{V_A}$ represents the
error due to the metamodel approximation. Nevertheless, $V_A$ is not a
linear transform of $Z_n(\X)$ and its full distribution cannot be
established.

\subsubsection{Variance estimates with Monte-Carlo integration}

To evaluate the Sobol' index $S_A$, it is possible to use the pick-freeze
approaches presented in Chapter 4 and in
\citet{sobol1993,sobol2007estimating,Jan12}. By considering the formula
given in \citet{sobol1993}, $S_A$ can be approximated by:
\begin{equation}\label{book4Chapter7_GP_4}
S_{A,N} = \frac{\frac{1}{N}\sum_{i=1}^N 
Z_n(\X^{(i)})Z_n(X_{~A}^{(i)}) - \left(\frac{1}{2N} 
\sum_{i=1}^N Z_n(\X^{(i)}) + Z_n(X_{~A}^{(i)})  \right)^2}
{
\frac{1}{N}\sum_{i=1}^N Z_n(\X^{(i)})^2 - 
\left(\frac{1}{2N} 
\sum_{i=1}^N Z_n(\X^{(i)}) + Z_n(X_{~A}^{(i)})  \right)^2
},
\end{equation}
where $(\X^{(i)},X_{~A}^{(i)})_{i=1,\dots,N}$ is a 
$N$-sample 
from the random variable  $(\X, \X^{\sim A})$. 

In particular, this approach avoids to compute the integrals presented in 
\citet{Oak04} and thus   
simplify the estimation of $V_A$ and $V$. Furthermore, it takes into account their joint 
distribution.

\paragraph{Remark.} This result can easily be extended to the total
Sobol' index $S^{\mbox{\scriptsize{tot}}}_i = \sum\limits_{A \supset \,i} S_A$.
The reader is referred to \citet{sobol2007estimating} and 
Variance-based sensitivity analysis: Theory and 
estimation algorithms in this handbook for
examples of pick-freeze estimates of $S_A^{\mbox{\scriptsize{tot}}}$.

\subsection{Numerical estimates of Sobol' indices by Gaussian process sampling}
\label{sec:7-03.4}

The sensitivity index $S_{A,N}$ (\ref{book4Chapter7_GP_4}) 
is 
obtained after 
substituting the Gaussian process $Z_n(\ve{x})$ for the original computational 
model $G(\Ve{x})$. Therefore, it is a random variable 
defined on the same probability space as $Z_n(\ve{x})$. 
The aim of this section is to present a simple methodology 
to get a sample  $S_{A,N}$ of $S_A$. From 
this sample, an estimate of $S_A$ (\ref{book4Chapter7_GP_5}) and a quantification of its 
uncertainty can be deduced.

\paragraph{Sampling from the Gaussian predictive distribution}

To obtain a realization of $S_{A,N}$, one has to obtain a 
sample of
$Z_n(\ve{x})$ on $(\X^{(i)},X_{~A}^{(i)})_{i=1,\dots,N}$ 
and then use
Eq.~(\ref{book4Chapter7_GP_4}). To deal with large $N$, an efficient
strategy is to sample $Z_n(\ve{x})$ using the Kriging conditioning
method, see for example  \citet{chiles1999geostatistics}.  
Consider first the
unconditioned, zero-mean Gaussian process:
\begin{equation}
  \tilde Z(\ve{x}) = \GP{0}{k(\ve{x},\ve{x}')}.
\end{equation}
Then, the Gaussian process:
\begin{equation}
  \tilde Z_n(\ve{x}) = m_n(\ve{x}) - \tilde m_n(\ve{x}) + \tilde Z(\ve{x}),
\end{equation}
where $\tilde m_n(\ve{x}) = \ff\tr(\ve{x}) \tilde{\bbeta} +
\rr\tr(\ve{x})\K^{-1} \left( \tilde Z(\cx) - \FF \tilde \bbeta\right)$
and $\tilde{\bbeta} = \left(\FF\tr \K^{-1} \FF \right)^{-1} \FF\tr
\K^{-1} \tilde Z(\cx)$ has the same distribution as $Z_n(\ve{x})$.
Therefore, one can compute realizations of $Z_n(\ve{x})$ from
realizations of $\tilde Z(\ve{x})$.  Since $\tilde Z(\ve{x})$ is not
conditioned, the problem is numerically easier. Among the available
Gaussian process sampling methods, several can be mentioned: Cholesky
decomposition \citep{R06}, Fourier spectral decomposition \citep{S99},
Karhunen-Loeve spectral decomposition \citep{R06} and the propagative
version of the Gibbs sampler \citep{lantuejoul2013simulation}.

\paragraph{Remark.} Let suppose that a new point $\ve{x}^{(n+1)}$ is
added to the experimental design $\cx$. A classical result of
conditional probability implies that the new predictive distribution
$[Z(\ve{x})|Z(\cx) = \cy, Z(\ve{x}^{(n+1)}) = G(\ve{x}^{(n+1)}),
\sigma^2, \ttheta]$ is identical to $[Z_n(\ve{x})| Z_n(\ve{x}^{(n+1)}) =
G(\ve{x}^{(n+1)}), \sigma^2, \ttheta]$.  Therefore, $Z_n(\ve{x})$ can be
viewed as an unconditioned Gaussian process and, using the Kriging
conditioning method, realizations of $[Z(\ve{x})|Z(\cx) = \cy,
Z(\ve{x}^{(n+1)}) = G(\ve{x}^{(n+1)}), \sigma^2, \ttheta]$ can be
derived from realizations of $Z_n(\ve{x})$ using the following equation:
\begin{equation}
Z_{n+1}(\ve{x}) = \frac{k_n(\ve{x}^{(n+1)}, \cx)}{k_n(\ve{x}^{(n+1)},\ve{x}^{(n+1)})}\left( 
G(\ve{x}^{(n+1)}) - Z_n(\ve{x}^{(n+1)}) \right) + Z_n(\ve{x}).
\end{equation}
Therefore, it is easy to calculate a new sample of 
$S_{A,N}$ after adding a new point $\ve{x}^{(n+1)}$ to the 
experimental design set $\cx$. This result is used in the R 
CRAN package ``sensitivity'' to perform sequential design   
for sensitivity analysis using a Stepwise Uncertainty 
Reduction (SUR) strategy \citep{Bec11}.

\subsubsection{Meta-model and Monte-Carlo sampling errors}

Let us denote by $\acc{S_{A,i}^N, ~i=1,\dots,m}$ a sample 
set of $S_{A,N}$
(\ref{book4Chapter7_GP_4}) where of size $m > 0$. From this sample set, the
following unbiased estimate of $S_A$ can be deduced:
\begin{equation}
  \label{SGPestimates}
  \hat S_A = \frac{1}{m} \sum_{i=1}^m S_{A,i}^N.
\end{equation}
with variance:
\begin{equation}\label{SGPUQ}
  \hat \sigma_{\hat S_A}^2 = \frac{1}{m-1} \sum_{i=1}^m \left( S_{A,i}^N - \hat S_A\right)^2.
\end{equation}
The term $\hat \sigma_{\hat S_A}^2$ represents the uncertainty on the
estimate of $S_A$ (\ref{book4Chapter7_GP_5}) due to the metamodel
approximation. Therefore, with the presented strategy, one can both
obtain an unbiased estimate of the sensitivity index $S_A$ and a
quantification of its uncertainty.

Finally it may be of interest to evaluate the error due to the
pick-freeze approximation and to compare it to the error due to the
metamodel.  To do so, one can use the central limit theorem
\citep{Jan12,ChastaingLeGratiet2015} or a bootstrap procedure
\citep{MultiFidelitySA}. In particular, a methodology to evaluate the
uncertainty on the sensitivity index due to both the Gaussian process
and to the pick-freeze approximations is presented in
\citet{MultiFidelitySA}.  It makes it possible to determine the value of
$N$ such that the pick-freeze approximation error is negligible compared
to that of the metamodel.

\subsection{Summary}
\label{sec:7-03.5}
Gaussian Process regression makes it possible to perform sensitivity
analysis on complex computational models using a limited number of model 
evaluations. An important feature of this method is that one can propagate the 
Gaussian process approximation error to the sensitivity index estimates.  
This allows the construction of sequential design strategies optimized for 
sensitivity analysis.  It also provides a powerful tool to
visualize the main effect of a group of variables and the uncertainty of
its estimate.  Another advantage of this approach is that Gaussian
process regression has been thoroughly investigated in the literature and can 
be used in various problems. For example, the method can be adapted for
non-stationary numerical models by using a treed Gaussian process
as in \citet{gramacy2012categorical}. Furthermore, it can also be used for
multifidelity computer codes, i.e.  codes which can be run at multiple
level of accuracy (see \citet{MultiFidelitySA}).

\section{Applications}
\label{sec:7-04}
In this section, metamodel-based sensitivity analysis is illustrated on
several academic and engineering examples.

\subsection{Ishigami function}
\label{sec:7-04.1}

The Ishigami function is given by:
\begin{equation}
G(x_1,x_2,x_3) = \sin (x_1) + 7  \sin(x_2)^2 + 0.1 x_3^4 \sin(x_1).
\end{equation}
The input distributions of $X_1, X_2$ and $X_3$ are uniform over the
interval $[-\pi, \pi]^3$.  This is a classical academic benchmark for
sensitivity analysis, with first-order Sobol' indices:
\begin{equation}
  S_1=0.3138 \qquad  S_2 = 0.4424 \qquad S_3  = 0.
\end{equation}
To compare polynomial chaos expansions and Gaussian process modeling on
this example, experimental designs of different sizes $n$ are
considered. For each size $n$, 100 Latin Hypercube Sampling sets (LHS) are
computed so as to replicate the procedure and assess statistical
uncertainty.

For the polynomial chaos approach, the coefficients are calculated based
on a degree-adaptive LARS strategy (for details, see
\citet{SudretJCP2011}), resulting in a sparse basis set. The maximum
polynomial degree is adaptively selected in the interval $3 \leq p \leq
15$ based on LOO cross-validation error estimates (see Eq.~(\ref{eq:723})).

For the Gaussian process approach, a tensorized Mat\'ern-$5/2$
covariance kernel is chosen (see \citet{R06}) with trend functions given
by:
\begin{equation}
 \ff\tr(\ve{x}) = \acc{1 ~~ x_2 ~~ x_2^2 ~~ x_1^3 ~~ x_2^3 ~~ x_1^4 ~~ x_2^4}.
\end{equation}
The hyper-parameters $\ttheta$ are estimated with a Leave-One-Out cross
validation procedure while the parameters $\bbeta$ and $\sigma^2$ are
estimated with a restricted maximum likelihood method.

\begin{figure}[!ht]
  \begin{center}
    \includegraphics[scale=0.45,angle=0]{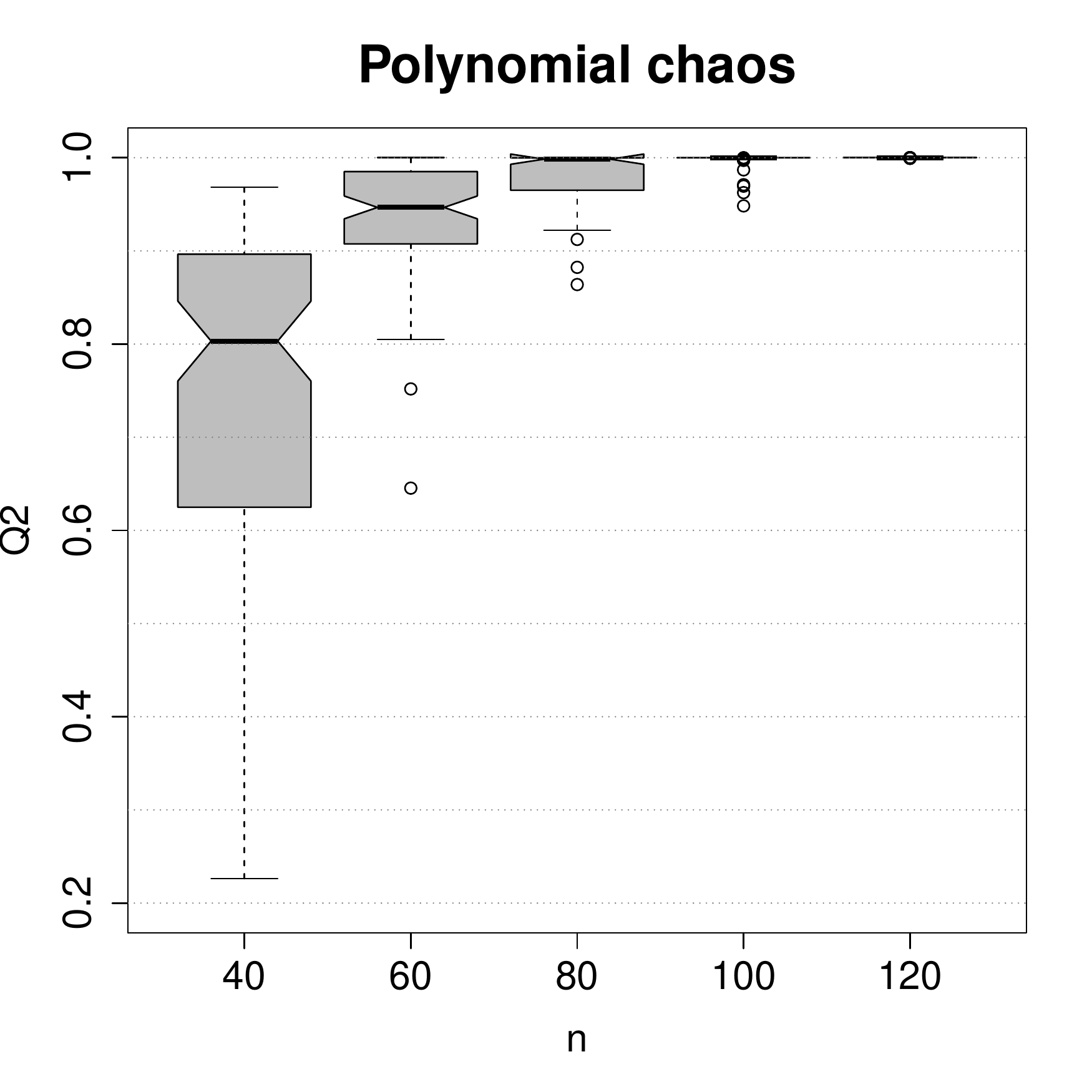}
    \includegraphics[scale=0.45,angle=0]{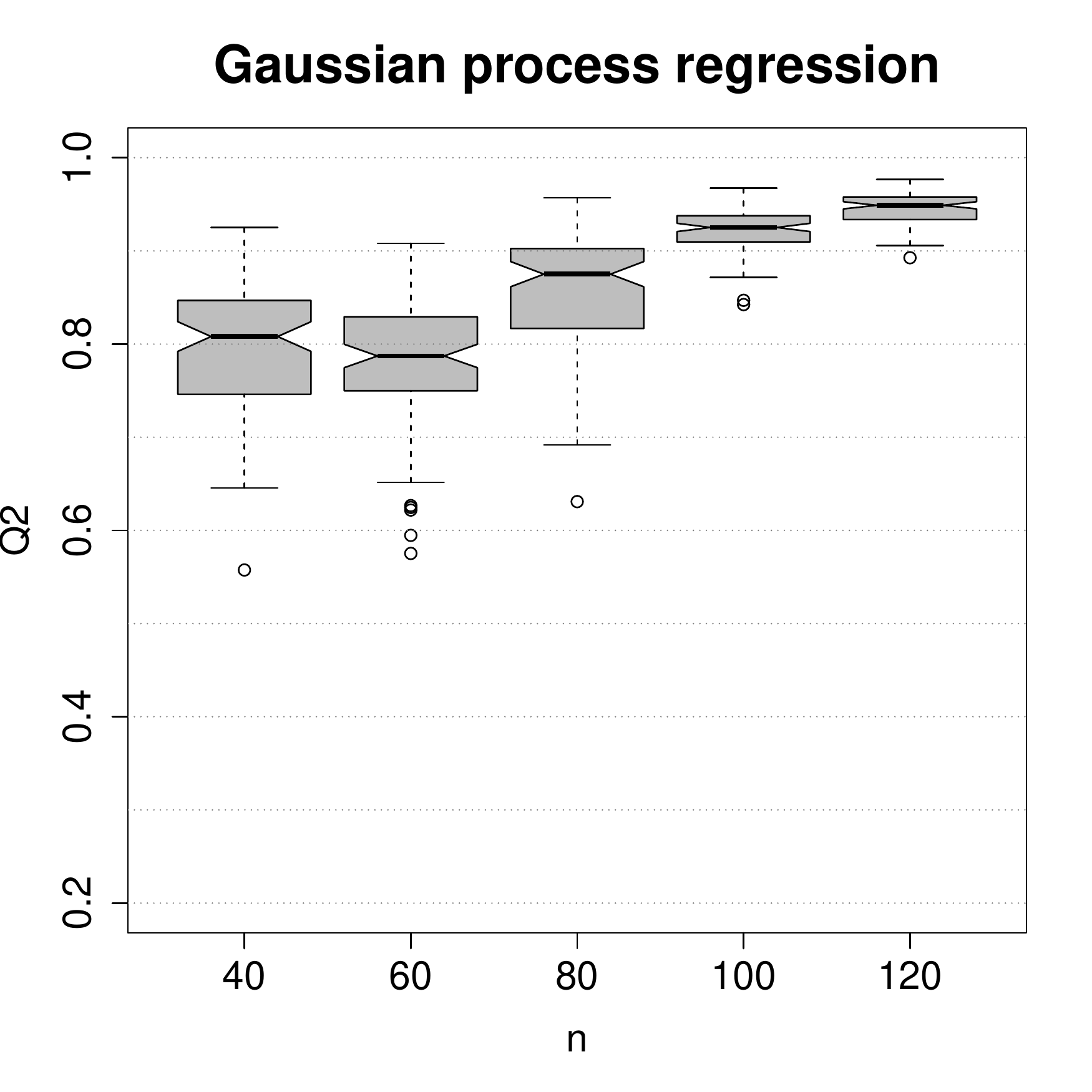}
    \caption{$Q^2$ coefficient as a function of the sample size $n$ for
      the Ishigami function. For each $n$, the box-plots represent the
      variations of $Q^2$ obtained over 100 LHS replications.}
    \label{E_metamodels}
  \end{center}
\end{figure}

First we illustrate in Figure \ref{E_metamodels} the accuracy of the
models with respect to the sample size $n$. The Nash-Sutcliffe model
efficiency coefficient (also called predictivity 
coefficient) is defined as follows:
\begin{equation}
Q^2 = 1 - \frac{\sum_{i=1}^{n_\mathrm{test}} 
(G(\ve{x}^{(i)}) - \hat G(\ve{x}^{(i)})  
)^2}{\sum_{i=1}^{n_\mathrm{test}} (G(\ve{x}^{(i)}) - \bar 
G   )^2}, \quad \bar G   = 
\frac{1}{{n_\mathrm{test}}} \sum_{i=1}^{n_\mathrm{test}}  
G(\ve{x}^{(i)}),
\end{equation}
where $\hat G(\ve{x}^{(i)})$ is the prediction given by the 
polynomial chaos
or the Gaussian process regression model on the $i^\mathrm{th}$ point of
a test sample of size ${n_\mathrm{test}} = 10,000$. This test sample set
is randomly generated from a uniform distribution. The closer $Q^2$ is
to 1, the more accurate the metamodel is.

We emphasize that checking the metamodel accuracy (see
Figure~\ref{E_metamodels}) is very important since a metamodel-based
sensitivity analysis provides sensitivity indices for the metamodel and
not for the true model $G(\ve{x})$. Therefore, the estimated indices are
relevant only if the considered surrogate model is accurate.

\begin{figure}[!ht]
  \begin{center}
    \includegraphics[scale=0.45,angle=0]{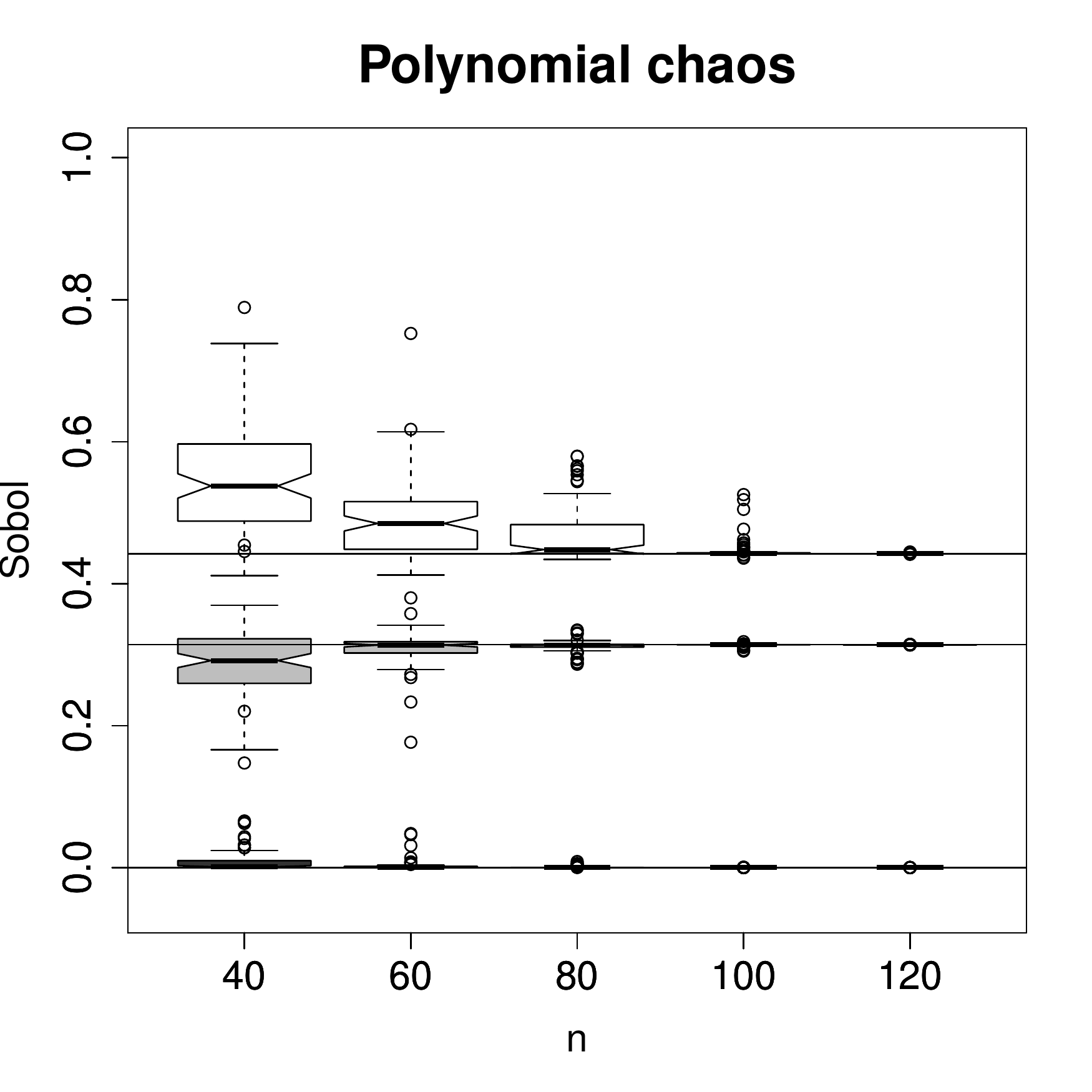}
    \includegraphics[scale=0.45,angle=0]{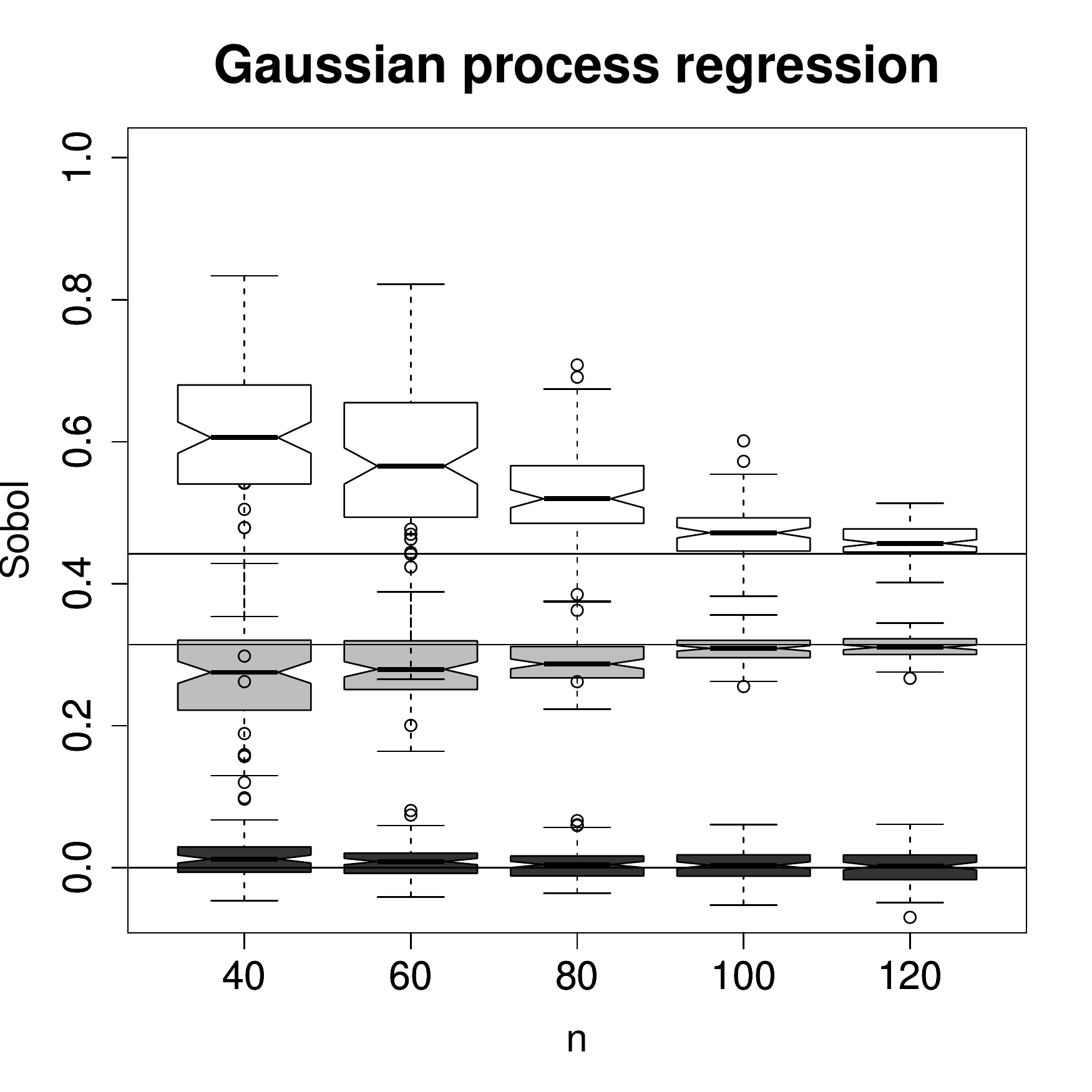}
    \caption{First-order Sobol' index estimates as a function of the
      sample size $n$ for the Ishigami function. The horizontal solid
      lines represent the exact values of $S_1$, $S_2$ and $S_3$. For
      each $n$, the box-plot represents the variations obtained from 100
      LHS replications. The validation set comprises $n_{test} = 10,000$
      samples.}    \label{S_metamodels}
  \end{center}
\end{figure}

Figure \ref{S_metamodels} shows the Sobol' index estimates with respect
to the sample size $n$.  For the Gaussian process regression approach,
the convergence for is reached for $n = 100$. It corresponds to a $Q^2$
coefficient greater than 90\%. Convergence of the PCE approach is
somewhat faster, with comparable accuracy achieved with $n = 60$ and
almost perfect accuracy for $n = 100$. Therefore, the convergence of the
estimators of the Sobol' indices in Eqs.~\eqref{eq:752} to
\eqref{eq:755} is expected to be comparable to that of $Q^2$. Note that
the PCE approach also provides second order- and total Sobol' indices
for free, as shown in \citet{SudretBookPhoon2015}.

\clearpage
\subsection{G-Sobol function}
\label{sec:7-04.2}

The G-Sobol function is given by :
\begin{equation}
G(\ve{x}) = \prod_{i=1}^d \frac{|4x_i - 2| + a_i}{1+a_i}, \quad a_i \geq 0.
\end{equation}
To benchmark the described metamodel-based sensitivity analysis methods
in higher dimension, we select $d=15$.  The exact first-order Sobol'
indices $S_i$ are given by the following equations:

\begin{equation}
  \begin{split}
    V_i &= \frac{1}{3(1+a_i)^2}, \quad i=1,\dots, d, \\
    V   &= \prod_{i=1}^d (1+V_i) -1, \\
    S_i &= V_i /V.
  \end{split}
\end{equation}
In this example, vector $\ve{a} = \acc{a_1,a_2,\dots,a_d}$ is equal to:
\begin{equation}
  \ve{a} = \acc{1, 2, 5, 10, 20, 50, 100, 500, 1000, 1000, 1000, 1000, 1000, 1000, 1000}.
\end{equation}

As in the previous section, different sample sizes $n$ are considered
and 100 LHS replications are computed for each $n$.  Sparse polynomial chaos 
expansions are obtained with the same strategy as for the Ishigami function: 
adaptive polynomial degree selection with $3 < p < 15$ and LARS-based 
calculation of the coefficients.
For the Gaussian process regression model, a tensorized Mat\'ern-5/2 covariance 
kernel is considered with a constant trend function $\ff(\ve{x}) = 1$. The
hyper-parameter $\ttheta$ is estimated with a Leave-One-Out cross
validation procedure and the parameters $\bbeta$ and $\sigma^2$ are
estimated with the maximum likelihood method.

\begin{figure}[!ht]
  \begin{center}
    \includegraphics[scale=0.45,angle=0]{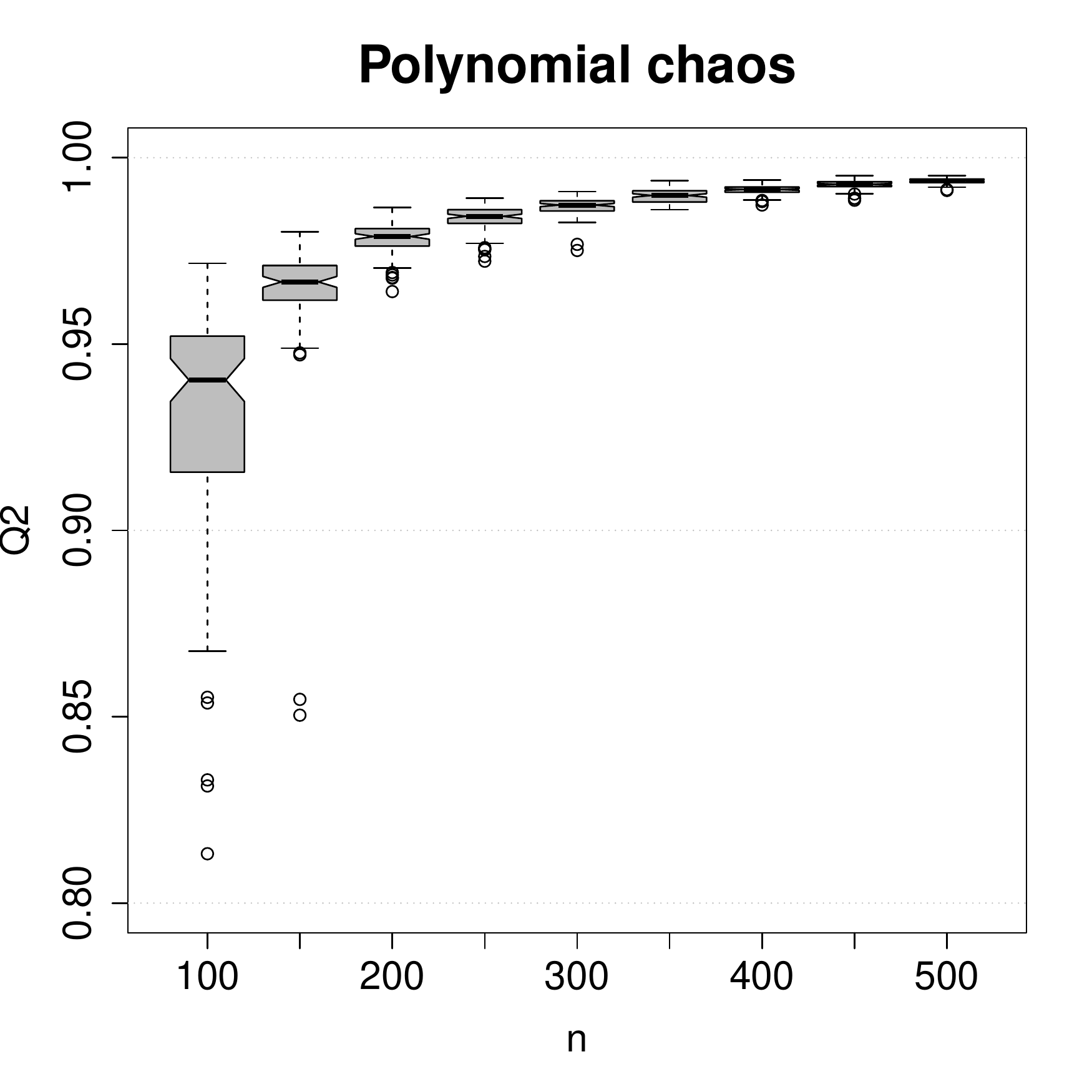}
    \includegraphics[scale=0.45,angle=0]{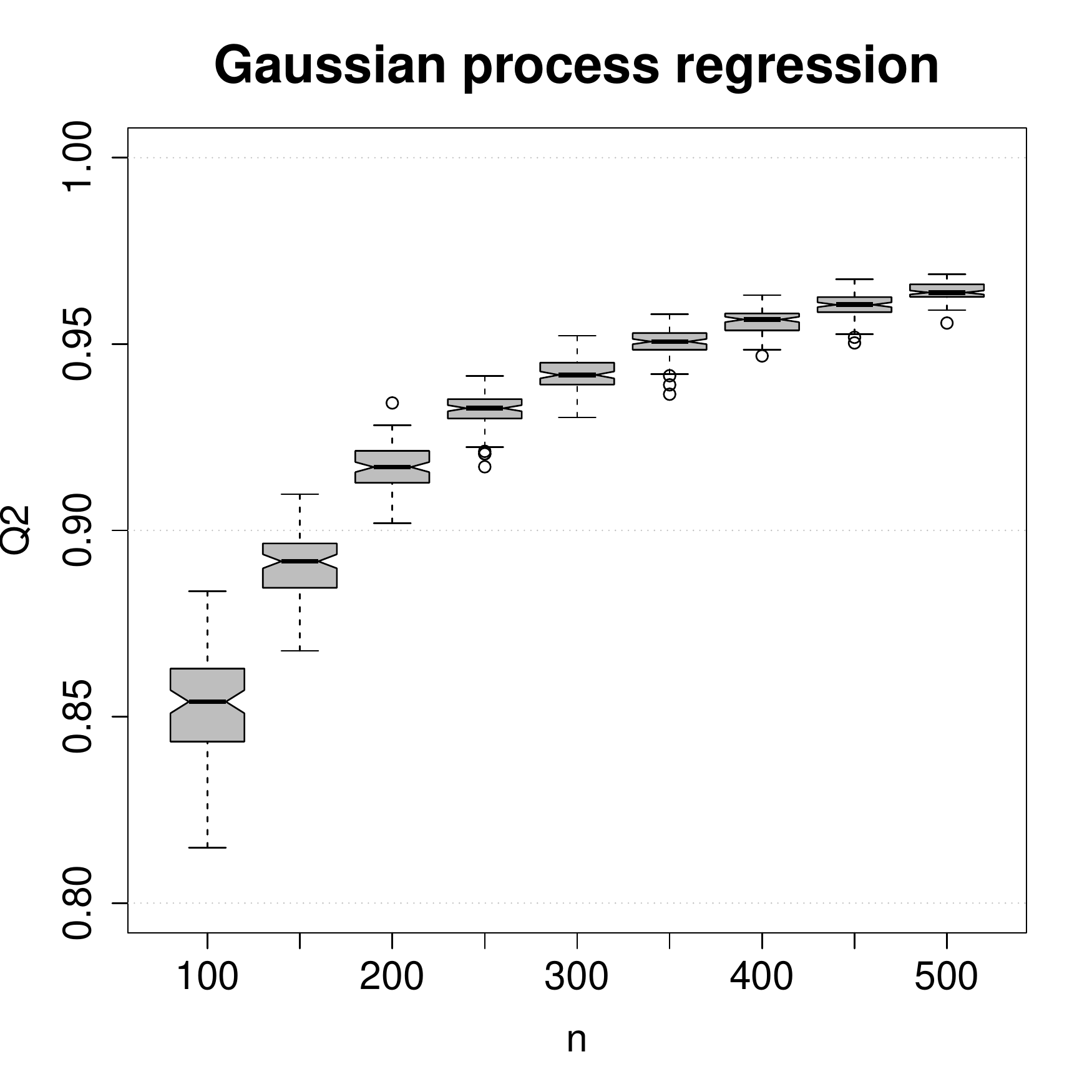}
    \caption{$Q^2$ coefficient as a function of the sample size $n$ for
      G-Sobol academic example. For each $n$, the box-plot represents the
      variations of $Q^2$ obtained from 100 LHS. 
      }  
      \label{E_metamodels_Gsobol}
  \end{center}
\end{figure}

\begin{figure}[h] 
  \begin{center}
    \includegraphics[scale=0.45,angle=0]{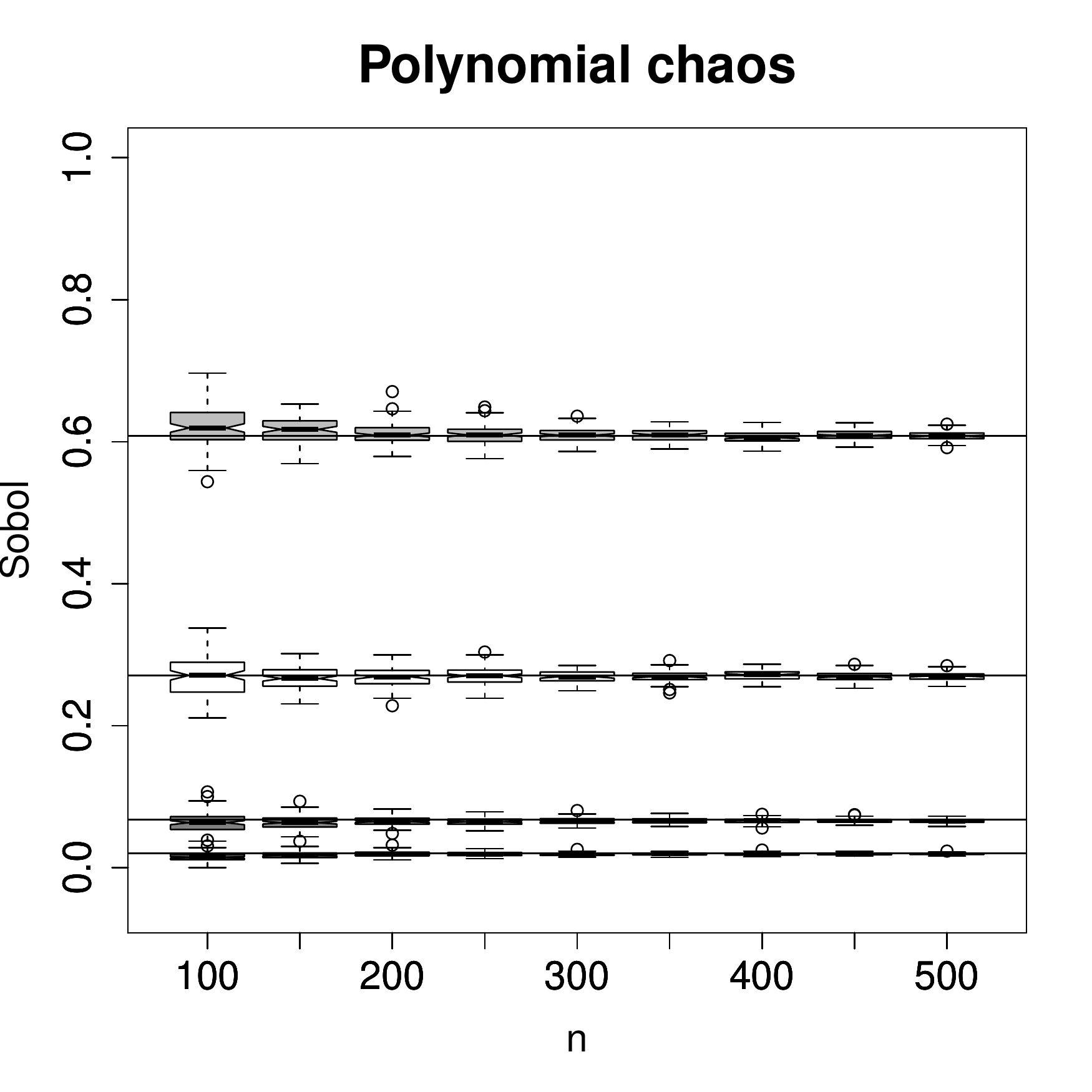}
    \includegraphics[scale=0.45,angle=0]{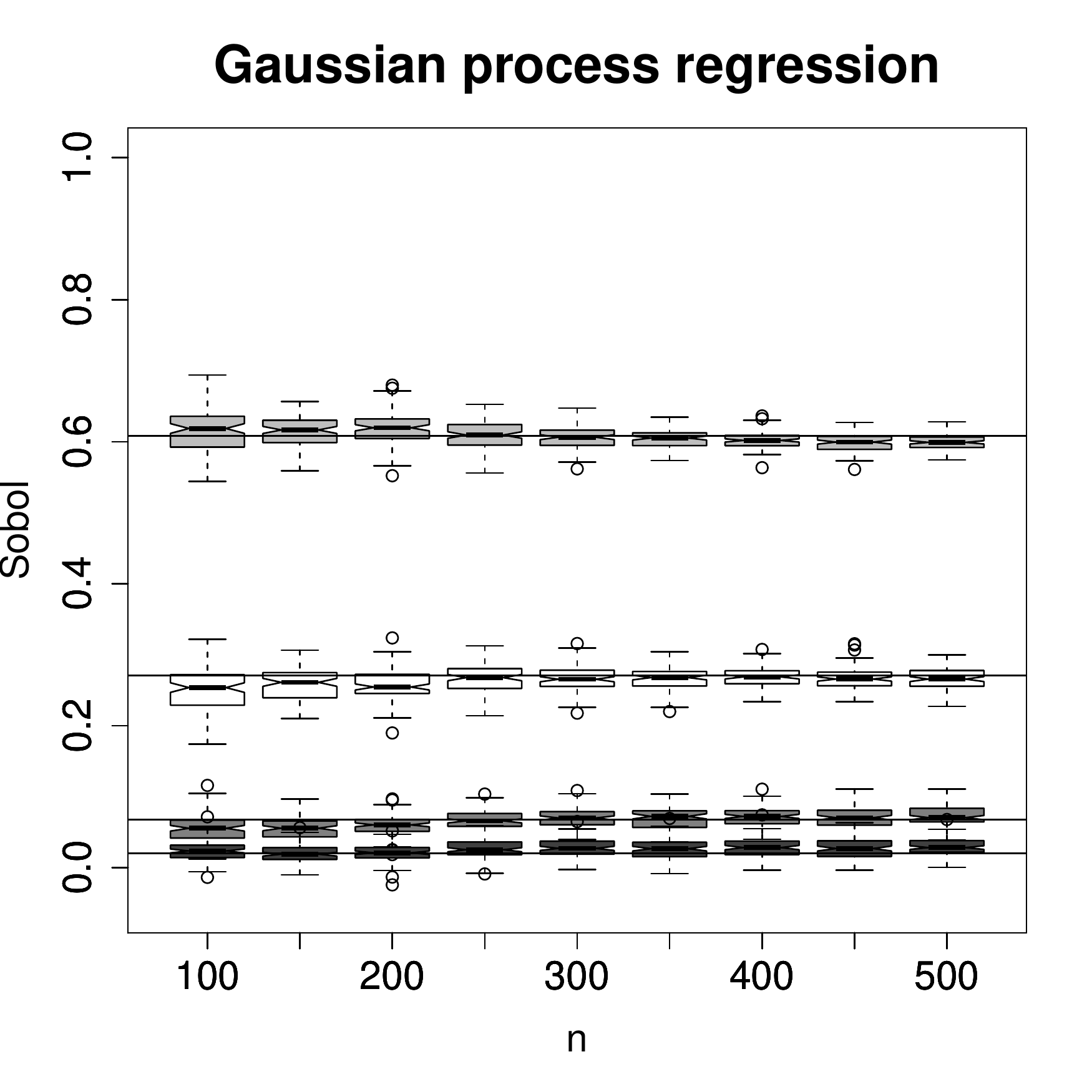}
    \caption{Sobol' index estimates with respect to the sample size $n$
      for G-Sobol function. The horizontal solid lines represent the
      true values of $S_1$, $S_2$, $S_3$ and $S_4$. For each $n$, the
      box-plot represents the variations obtained from 100 LHS. }
    \label{S_metamodels_Gsobol}
  \end{center}
\end{figure}

The accuracy of the metamodels with respect to $n$ is presented in
Figure~\ref{E_metamodels_Gsobol}. It is computed from a test sample set of
size $n_\mathrm{test} = 10,000$.  The convergence of the estimates of
the first four first-order Sobol' indices is represented in Figure
\ref{S_metamodels_Gsobol}. Both metamodel-based estimations yield
excellent results already with $n = 100$ samples in the experimental
design. This is expected due to the good accuracy of both metamodels for
all the $n$ considered (see Figure~\ref{E_metamodels_Gsobol}).

Finally, Table~\ref{tab:SGSobol} provides the Sobol' index estimates
median and root mean square error for $n=100$ and $n=500$.  As presented
in Figure \ref{S_metamodels_Gsobol}, the estimates of the largest Sobol' 
indices are very accurate. Note that the remaining first order
indices are insignificant. One can observe that the RMS error over the
100 LHS replications is slightly smaller when using PCE for both $n=100$
and $n=500$ ED points. Note that the second order- and total Sobol'
indices are also available for free when using PCE. 

\begin{table}[!ht]
  \caption{Sobol' index estimates for the G-Sobol function. The median
    and the root mean square error  
    (RMSE) of the estimates are given for $n=100$ and
    $n=500$.} \label{tab:SGSobol}
%
  \begin{center}
    \begin{tabular}{|c|c|c|c|c|c|c|c|c|c|}
    	\hline
    	\multicolumn{2}{|c|}{ }  & \multicolumn{4}{|c|}{Polynomial chaos 
    	expansion}                         & \multicolumn{4}{|c|}{Gaussian 
    	process regression}                               \\ \hline
    	\multicolumn{2}{|c|}{ }  & \multicolumn{2}{|c|}{Median} & 
    	\multicolumn{2}{|c|}{RMSE}                & 
    	\multicolumn{2}{|c|}{Median}             & \multicolumn{2}{|c|}{RMSE}\\ 
    	\hline
    	 Index   &    Value      &  100  &         500          &         
    	 100         &         500         &        100         &         
    	 500         &  100  &        500        \\ \hline
    	 $S_1$   &    0.604      & 0.619 &        0.607         &        
    	 0.034        &        0.007        &       0.618        &        
    	 0.599        & 0.035 &       0.012       \\
    	 $S_2$   &    0.268      & 0.270 &        0.269         &        
    	 0.027        &        0.005        &       0.233        &        
    	 0.245        & 0.046 &       0.026       \\
    	 $S_3$   &    0.067      & 0.063 &        0.065         &        
    	 0.014        &        0.003        &       0.045        &        
    	 0.070        & 0.029 &       0.016       \\
    	 $S_4$   &    0.020      & 0.014 &        0.019         &        
    	 0.008        &        0.001        &       0.008        &        
    	 0.023        & 0.018 &       0.013       \\
    	 $S_5$   &    0.005      & 0.002 &        0.005         &        
    	 0.003        &        0.001        & 8.6$\times10^{-4}$ & 
    	 1.8$\times10^{-3}$  & 0.014 &       0.013       \\
    	 $S_6$   &    0.001      & 0.000 &  7.2$\times10^{-4}$  &        
    	 0.001        & 3.5$\times10^{-4}$  & 6.4$\times10^{-4}$ & 
    	 5.3$\times10^{-4}$  & 0.013 &       0.013       \\
    	 $S_7$   &    0.000      & 0.000 &  1.1$\times10^{-4}$  & 
    	 1.1$\times10^{-3}$  & 1.4$\times10^{-4}$  & 5.3$\times10^{-4}$ & 
    	 3.0$\times10^{-4}$  & 0.013 &       0.013       \\
    	 $S_8$   &    0.000      & 0.000 &        0.000         & 
    	 3.3$\times10^{-4}$  & 1.7$\times10^{-5}$  & 6.5$\times10^{-4}$ & 
    	 7.1$\times10^{-4}$  & 0.013 &       0.013       \\
    	 $S_9$   &    0.000      & 0.000 &        0.000         & 
    	 4.1$\times10^{-4}$  & 1.1$\times10^{-5}$  & 8.5$\times10^{-4}$ & 
    	 4.4$\times10^{-4}$  & 0.14  &       0.013       \\
    	$S_{10}$ &    0.000      & 0.000 &        0.000         & 
    	2.4$\times10^{-4}$  & 1.1$\times10^{-5}$  & 2.2$\times10^{-4}$ & 
    	1.7$\times10^{-4}$  & 0.013 &       0.013       \\
    	$S_{11}$ &    0.000      & 0.000 &        0.000         & 9.5 
    	$\times10^{-4}$ & 1.2$\times10^{-5}$  & 5.5$\times10^{-4}$ & 
    	-9.9$\times10^{-5}$ & 0.013 &       0.013       \\
    	$S_{12}$ &    0.000      & 0.000 &        0.000         & 
    	5.2$\times10^{-4}$  & 2.1 $\times10^{-5}$ & 2.6$\times10^{-4}$ & 
    	4.1$\times10^{-4}$  & 0.013 &       0.013       \\
    	$S_{13}$ &    0.000      & 0.000 &        0.000         & 
    	5.1$\times10^{-4}$  & 5.9 $\times10^{-6}$ & 9.8$\times10^{-4}$ & 
    	4.7$\times10^{-4}$  & 0.013 &       0.013       \\
    	$S_{14}$ &    0.000      & 0.000 &        0.000         & 
    	8.8$\times10^{-4}$  & 1.9 $\times10^{-5}$ & 1.8$\times10^{-4}$ & 
    	6.9$\times10^{-4}$  & 0.013 &       0.013       \\
    	$S_{15}$ &    0.000      & 0.000 &        0.000         & 
    	8.6$\times10^{-4}$  & 9.7$\times10^{-6}$  & 7.2$\times10^{-4}$ & 
    	3.1$\times10^{-4}$  & 0.013 &       0.013       \\ \hline
    \end{tabular}
  \end{center}
\end{table}

\clearpage
\subsection{Morris function}
\label{sec:7-04.5}
The Morris function is given by \citep{Schoebi2015}: 

\begin{equation}\label{eq:MorrisFunction}
G(\ve{x}) = \sum_{i=1}^{20} \beta_i w_i + 
\sum_{i<j}^{20} \beta_{ij}w_i w_j + 
\sum_{i<j<l}^{20}
\beta_{ijl}w_i w_j w_l 
+ 5 w_1 w_2 w_3 w_4
\end{equation}
where $X_i \sim \cu(0,1), \, i=1,\ldots,20 $ and $w_i = 2(x_i -1/2)$ for all $i$ except for $i=3,5,7$ where $w_i = 2(1.1 x_i/(x_i+0.1) - 1/2)$. The coefficients are defined as $\beta_i = 20,\, i=1, \ldots, 10$; $\beta_{ij} = -15, \, i,j = 1, \ldots, 6$; $\beta_{ijl} = -10, \, i,j,l = 1, \ldots, 5$. The remaining coefficients are set equal to $\beta_i = (-1)^i$ and $\beta_{ij} = (-1)^{i+j}$ and all the rest are zero. The reference values of the first-order Sobol' indices of the Morris function are calculated by a large Monte Carlo-based sensitivity analysis ($n = 10^6$).
 
As in the previous section different sample sizes $n$ are considered and 100 
LHS replications are computed for each $n$. Sparse polynomial chaos expansions 
are obtained by adaptive polynomial degree selection $5<p<13$ and LARS-based 
calculation of the coefficients.

\begin{figure}[!ht]
	\begin{center}
		\includegraphics[scale=0.45,angle=0]{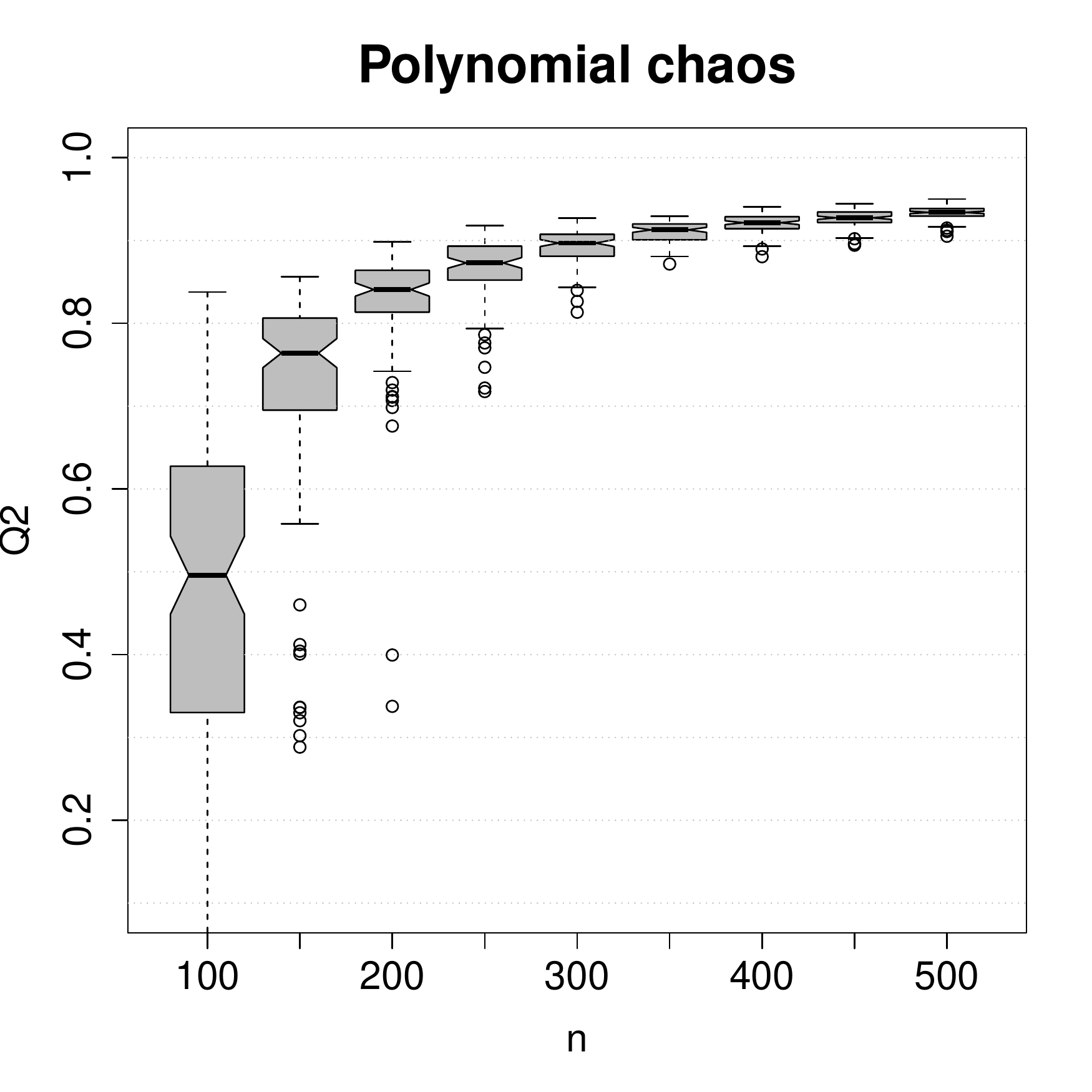} 
		\includegraphics[scale=0.45,angle=0]{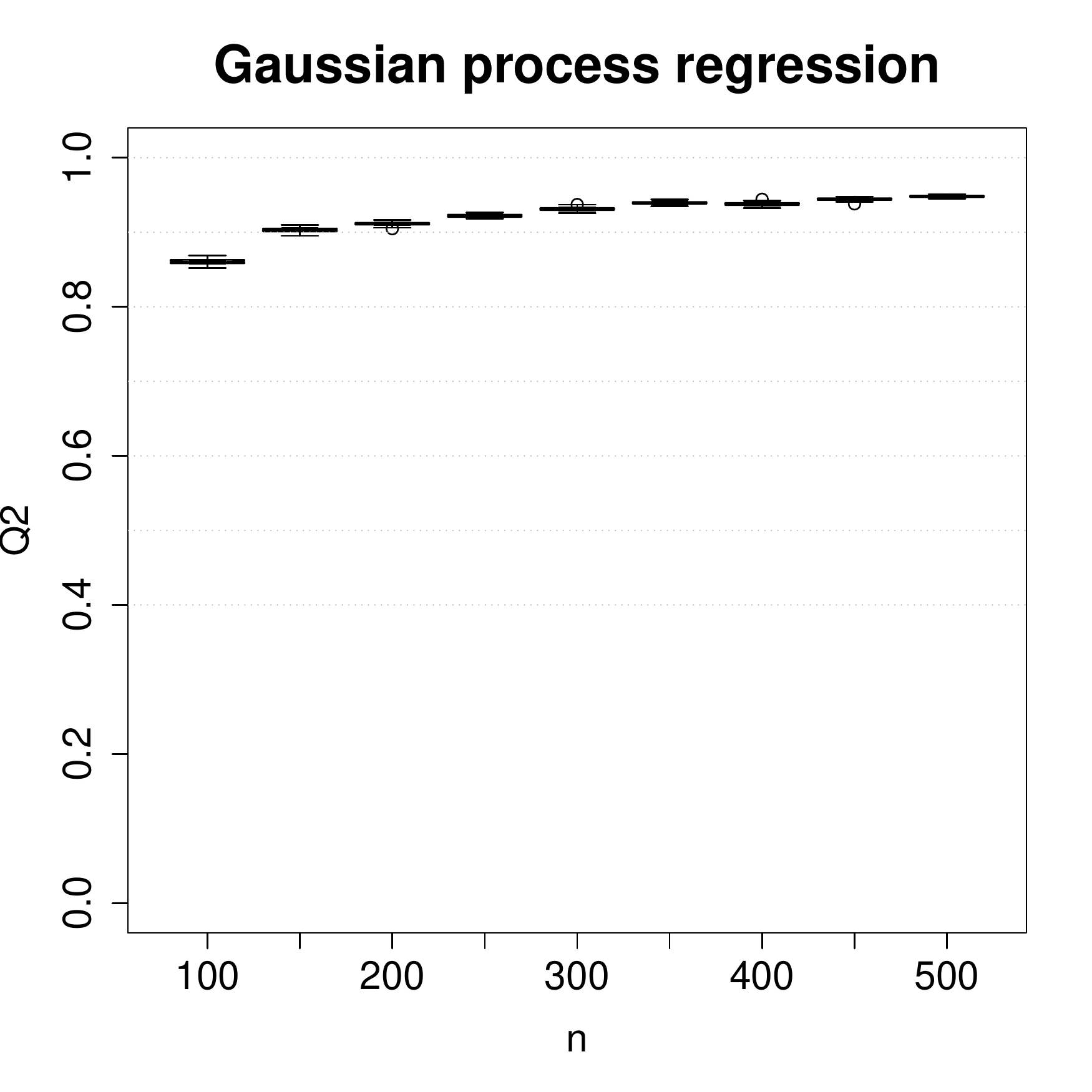}  
		\caption{$Q^2$ coefficient as a function of the 
		sample size $n$ for the Morris function example. 
		For each $n$, the box-plot represents the 
		variations of $Q^2$ obtained from 100 LHS. 
		}
		\label{fig:Q2_Morris}
	\end{center}
\end{figure}

The accuracy of the metamodels with respect to $n$ is presented in 
Figure~\ref{fig:Q2_Morris}. It is computed from a test sample of size $n_{test} 
= 10,000$. As expected due to the complexity and dimensionality of the Morris 
function, both metamodels show a slower overall convergence rate with the 
number of samples with respect to the previous examples. 
Polynomial chaos expansions show in this case remarkably more scattering in 
their performance for smaller experimental designs with respect to Gaussian 
process regression. This is likely due to the comparatively large amount of 
prior information in the form of trend functions provided to the Gaussian 
process, not used for PCE.

\begin{figure}[!ht]
	\begin{center}
		\includegraphics[scale=0.45,angle=0]{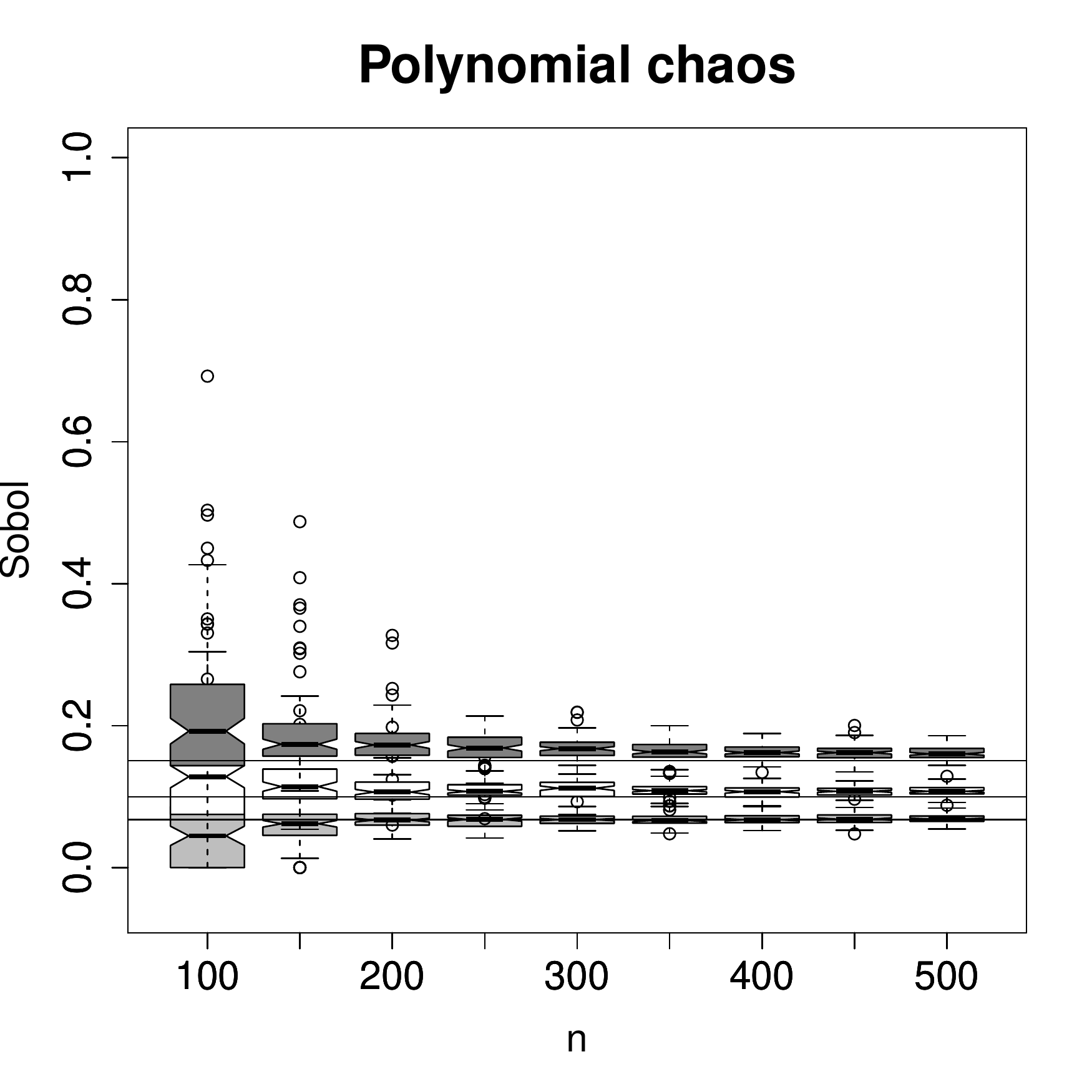}
		\includegraphics[scale=0.45,angle=0]{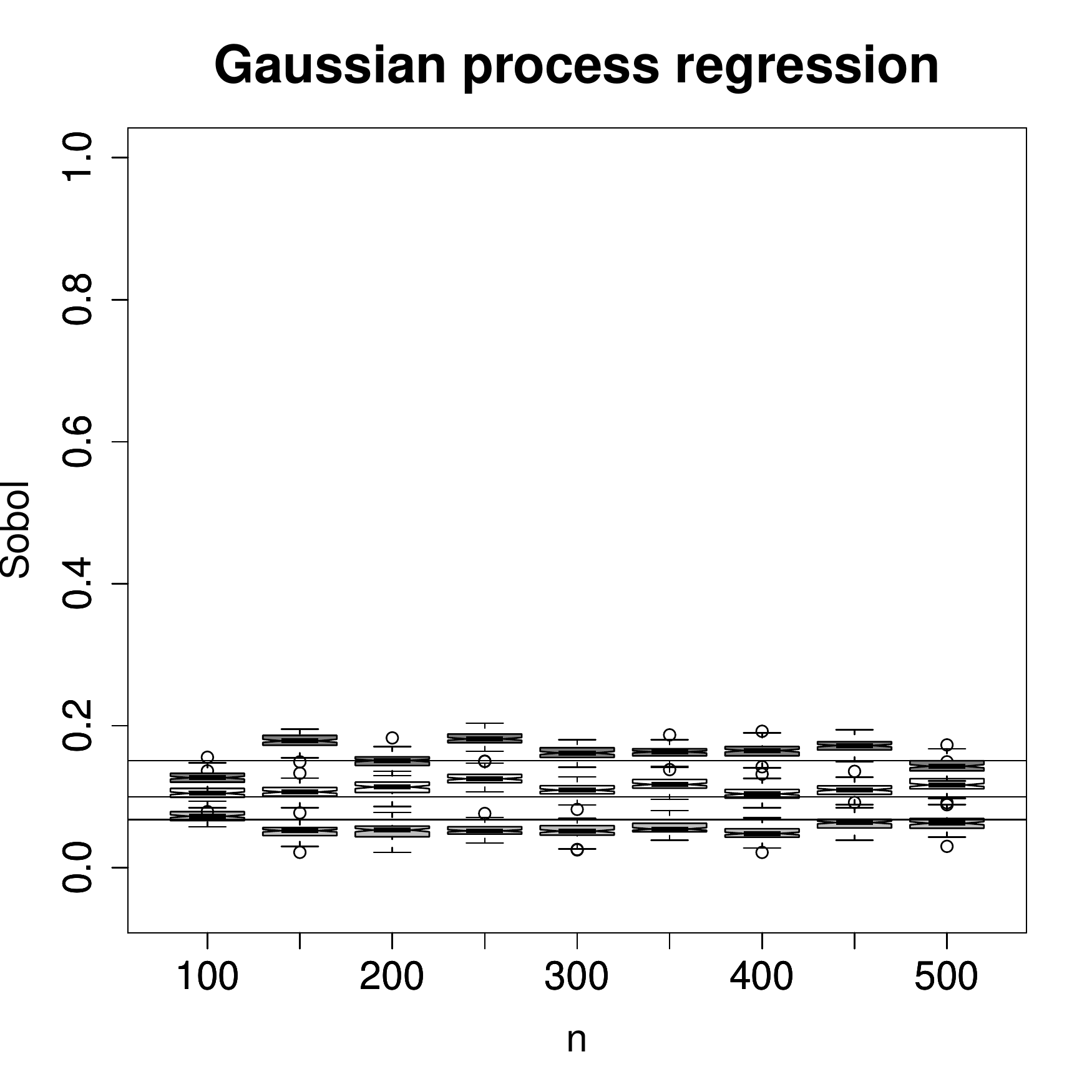}
		\caption{First-order Sobol' index estimates as a 
		function of the sample size $n$ for the Morris 
		function. The horizontal solid lines represent the 
		exact values of $S_3$, $S_8$ and $S_9$. For each 
		$n$, the box-plot represents the variations 
		obtained from 100 LHS replications. 
		}
		\label{fig:Sobol_Morris}
	\end{center}
\end{figure}

The convergence of the estimates of three selected first-order 
Sobol' indices (the largest $S_9$ and two intermediate ones $S_3$ and $S_8$) is 
represented in Figure~\ref{fig:Sobol_Morris}. Both methods perform very well 
with as few as 250 samples. PCE, however, shows a more standard convergence 
behavior both in mean value in dispersion. Gaussian process regression 
retrieves the Sobol' estimates very accurately even with extremely small 
experimental designs, but no clear convergence pattern can be seen for larger 
datasets.

Finally, Table~\ref{tbl:Sobol_Morris} provides the a detailed breakdown of the 
Sobol' index estimates, including median and root mean square error (RMSE), for 
$n=100$ and $n=500$. 

\begin{table}[!ht]
	\caption{First-order Sobol' indices estimation for the Morris function. The median and the root mean square error (RMSE) of the estimates are given for $n=100$ and $n=500$.}
	\label{tbl:Sobol_Morris}
	\vspace{0.3cm}
	
	\begin{tabular}{|c|c|c|c|c|c|c|c|c|c|}
		\hline
		\multicolumn{2}{|c|}{} & \multicolumn{4}{|c|}{Polynomial chaos expansion}              & \multicolumn{4}{|c|}{Gaussian process regression} \\ \hline
		\multicolumn{2}{|c|}{} & \multicolumn{2}{|c|}{Median} & \multicolumn{2}{|c|}{RMSE}	& \multicolumn{2}{|c|}{Median} & \multicolumn{2}{|c|}{RMSE}	 \\ \hline
Index 		& Value 	& 100 	     & 500 	  & 100        & 500 	    & 100   	& 500 		& 100 		& 500   	\\ \hline
$S_2$		&  0.005	&  0.000    &  0.005    &  0.252    &  0.017    &	0.011    &	0.004  &	0.109	&	0.085	\\
$S_3$		&  0.008	&  0.000    &  0.009    &  0.175    &  0.027    &	0.006    &	0.007  &	0.089	&	0.088	\\
$S_1$		&  0.017	&  0.000    &  0.015    &  0.304    &  0.047    &	0.003    &	0.017  &	0.130	&	0.109	\\
$S_4$		&  0.009	&  0.000    &  0.009    &  0.119    &  0.023    &	0.017    &	0.011  &	0.130	&	0.097	\\
$S_5$		&  0.016	&  0.000    &  0.015    &  0.230    &  0.043    &	0.005    &	0.016  &	0.120	&	0.109	\\
$S_6$		&  0.000	&  0.000    &  0.000    &  0.061    &  0.003    &	0.000    &	0.000  &	0.061	&	0.070	\\
$S_7$		&  0.069	&  0.045    &  0.068    &  0.585    &  0.058    &	0.072    &	0.062  &	0.095	&	0.123	\\
$S_8$		&  0.100	&  0.128    &  0.107    &  0.950    &  0.105    &	0.105    &	0.116  &	0.108	&	0.211	\\
$S_9$		&  0.150	&  0.192    &  0.160    &  1.241    &  0.143    &	0.127    &	0.143  &	0.246	&	0.117	\\
$S_{10}$	&  0.100	&  0.133    &  0.106    &  0.875    &  0.092    &	0.138    &	0.111  &	0.404	&	0.155	\\
$S_{11}$	&  0.000	&  0.000    &  0.000    &  0.185    &  0.003    &	0.004    &	0.000  &	0.088	&	0.074	\\
$S_{12}$	&  0.000	&  0.000    &  0.000    &  0.083    &  0.004    &	0.000    &	0.000  &	0.064	&	0.077	\\
$S_{13}$	&  0.000	&  0.000    &  0.000    &  0.081    &  0.003    &	0.000    &	0.000  &	0.064	&	0.074	\\
$S_{14}$	&  0.000	&  0.000    &  0.000    &  0.020    &  0.003    &	0.000    &	0.000  &	0.070	&	0.078	\\
$S_{15}$	&  0.000	&  0.000    &  0.000    &  0.140    &  0.003    &	0.000    &	0.000  &	0.065	&	0.075	\\
$S_{16}$	&  0.000	&  0.000    &  0.000    &  0.040    &  0.005    &	0.001    &	0.000  &	0.077	&	0.074	\\
$S_{17}$	&  0.000	&  0.000    &  0.000    &  0.264    &  0.004    &	0.000    &	0.000  &	0.065	&	0.075	\\ 		
$S_{18}$	&  0.000	&  0.000    &  0.000    &  0.084    &  0.004    &	0.000    &	0.000  &	0.064	&	0.075	\\
$S_{19}$	&  0.000	&  0.000    &  0.000    &  0.083    &  0.004    &	0.000    &	0.000  &	0.064	&	0.076	\\
$S_{20}$	&  0.000	&  0.000    &  0.000    &  0.049    &  0.004    &	0.000    &	0.000  &	0.064	&	0.075	\\
		\hline
	\end{tabular}

\end{table}

\clearpage
\subsection{Maximum deflection of a truss structure}
\label{sec:7-04.3}
\begin{figure}[!ht]
  \begin{center}
    \includegraphics[width=.7\textwidth, trim=50 200 50 190, clip =
    true]{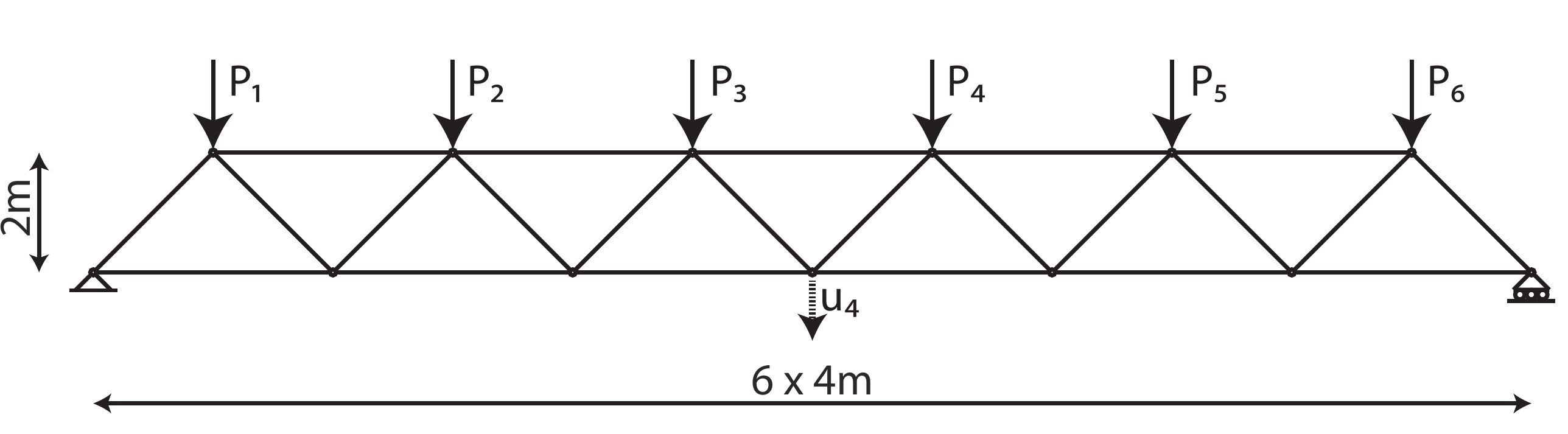} 
    \caption{Model of a truss structure with 23 members. The quantity of interest is the maximum 
      displacement at mid-span $u_4$.}
    \label{TrussStructure}
  \end{center}
\end{figure}

Sensitivity analysis is also of great interest for engineering models
whose input parameters may have different distributions. As an example
consider the elastic truss structure depicted in
Figure~\ref{TrussStructure} (see \eg \citet{BlatmanCras2008}). This
truss is made of two types of bars, namely horizontal bars with
cross-section $A_1$ and Young's modulus (stiffness) $E_1$ on the one
hand oblique bars with cross-section $A_2$ and Young's modulus
(stiffness) $E_2$ on the other hand. The truss is loaded with six
vertical loads applied on the top chord. Of interest is the maximum
vertical displacement (called deflection) at mid-span. This quantity is
computed using a finite element model comprising elastic bar elements.

The various parameters describing the behavior of 
this truss structure are modeled by independent random variables that 
account for the uncertainty in both the physical properties of the structure 
and the applied loads. Their distributions are gathered in
Table~\ref{tab:TrussInput}.

\begin{table}[!ht] \centering
  \caption{Probabilistic input model of the truss structure}
  \label{tab:TrussInput}
  \begin{tabular}{llcc}
    \hline
    Variable & 
    Distribution & 
    Mean & 
    Standard Deviation\\
    
    $E_1$, $E_2$ (Pa) &
    Lognormal &
    $2.1\times10^{11}$ &
    $2.1\times10^{10}$\\
    
    $A_1$ (m\textsuperscript{2}) &
    Lognormal &
    $2.0\times10^{-3}$ &
    $2.0\times10^{-4}$\\
    
    $A_2$ (m\textsuperscript{2}) &
    Lognormal &
    $1.0\times10^{-3}$ &
    $1.0\times10^{-4}$\\
    
    $P_1$-$P_6$ (N) &
    Gumbel &
    $5.0\times10^{4}$ &
    $7.5\times10^3$\\
    \hline
  \end{tabular}
\end{table}

These input variables are collected in the random vector
\begin{equation}
  \ve{X} = \acc{E_1,E_2,A_1,A_2, P_1,\dots,P_6}.
\end{equation}
Using this notation, the maximal deflection of interest is cast as:
\begin{equation}
  u_4= \cm^{\text{FE}}(\Ve{X} ).
\end{equation}

Different sparse polynomial chaos expansions are calculated assuming a maximal
degree $3<p<10$ using LARS and the best expansion (in terms of smallest
LOO error) is retained.
For the Gaussian process regression model, a tensorized Mat\'ern-5/2
covariance kernel is considered with a constant trend function
$\ff(\ve{x}) = 1$. The hyper-parameter $\ttheta$ is estimated with a
Leave-One-Out cross validation procedure and the parameters $\bbeta$ and
$\sigma^2$ are estimated with the maximum likelihood method.  

The first order Sobol' indices obtained from PCE and GP metamodels are
reported in Table~\ref{tab:TrussSobolIndices} in the case when the
experimental design is of size 100. In decreasing importance order, the
important variables are the properties of the chords (horizontal bars),
then the loads close to mid-span, namely $P_3$ and $P_4$. The Sobol'
indices of the latter are identical due to the symmetry of the model.
Then come the loads $P_2$ and $P_5$.  The other variables (the loads
$P_1$ and $P_6$ and the properties of the oblique bars) appear
unimportant.

\begin{table}[!ht] \centering
  \caption{Truss structure -- First order Sobol' indices} \label{tab:TrussSobolIndices} 
  \begin{tabular}{cccc}
    \hline
    Variable & Reference &  PCE       &   Gaussian Process   \\\hline
    $A_1$    & 0.365     &  0.366     &    0.384             \\
    $E_1$    & 0.365     &  0.369     &    0.362             \\
    $P_3$    & 0.075     &  0.078     &    0.075             \\
    $P_4$    & 0.074     &  0.076     &    0.069             \\
    $P_5$    & 0.035     &  0.036     &    0.029             \\
    $P_2$    & 0.035     &  0.036     &    0.028             \\
    $A_2$    & 0.011     &  0.012     &    0.015             \\
    $E_2$    & 0.011     &  0.012     &    0.008             \\
    $P_6$    & 0.003     &  0.005     &    0.002             \\
    $P_1$    & 0.002     &  0.005     &    0.000             \\\hline
  \end{tabular}
\end{table}

\begin{figure}[!ht]
  \begin{center}
    \includegraphics[scale=0.45,angle=0]{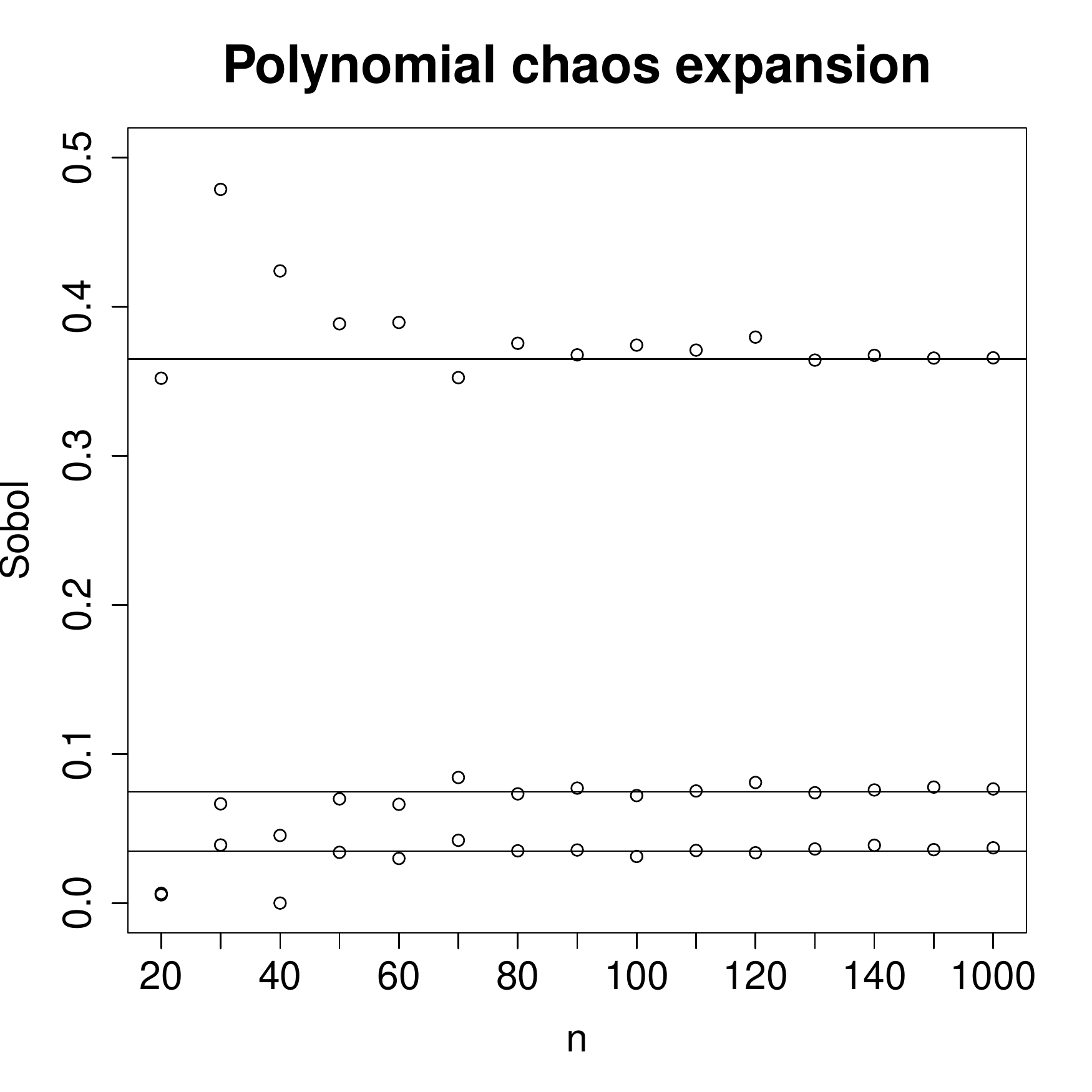}
    \includegraphics[scale=0.45,angle=0]{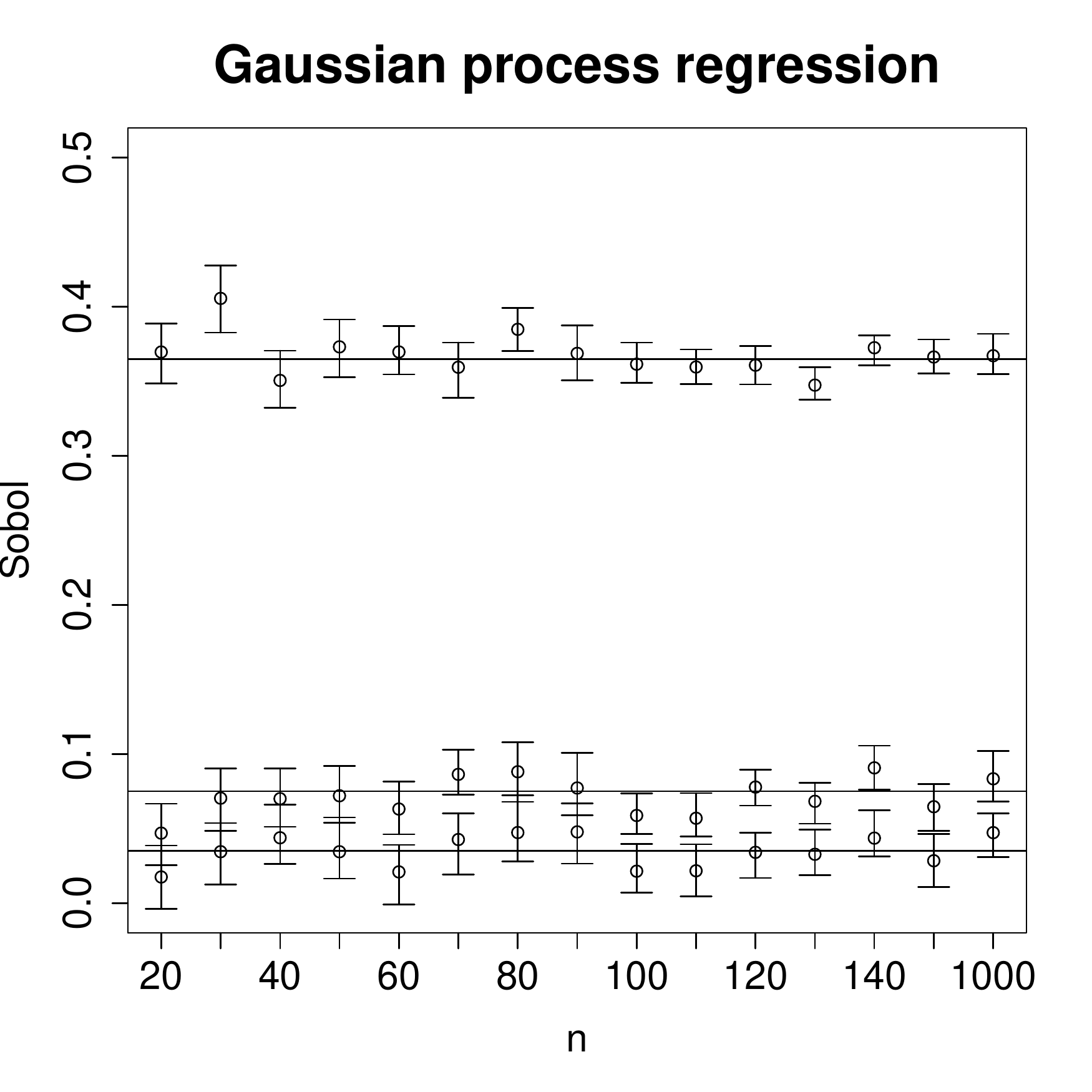}
    \caption{Truss structure -- First-order Sobol' index estimates as a
      function of the sample size $n$ for the truss model. The
      horizontal solid lines represent reference values of input
      variables $E_1$, $P_3$ and $P_5$ from a Monte Carlo estimate on
      6,000,000 samples.}     \label{S_metamodels_Truss}
  \end{center}
\end{figure}

The estimates of the three largest first-order Sobol' indices which
correspond to variables $E_1$, $P_3$ and $P_5$ obtained for various
sizes $n$ of the LHS experimental design are plotted in
Figure~\ref{S_metamodels_Truss} as a function of $n$. The reference solution is 
obtained by Monte-Carlo sampling with a sample set of size 6,000,000. Both PCE- and
GP-based Sobol' indices converge to stable estimates as soon as $n \ge
60$.

\section{Conclusions}
Sobol' indices are recognized as good descriptors of the 
sensitivity of the output of a computational model to its 
various input parameters. Classical estimation methods 
based on Monte Carlo simulation are computationally 
expensive though. The required costs, in the order of $10^3 
- 10^4$ model evaluations, are often not compatible with 
the advanced simulation models encountered in engineering 
applications.

For this reason, surrogate models may be first built up from a limited
number of runs of the computational model (the so-called experimental
design), and the sensitivity analysis is then carried out by
substituting the surrogate model for the original one.

Polynomial chaos expansions and Gaussian processes are two popular
methods that can be used for this purpose. The advantage of the PCE
approach is that the Sobol' indices at any order may be computed
{\em analytically} once the expansion is available. In this
contribution, least-square minimization techniques are presented to
compute the PCE coefficients, yet any intrusive or non intrusive method
could be used as an alternative.

In contrast Gaussian process surrogate models are used together with
Monte Carlo simulation for estimating the Sobol' indices. The advantage
of this approach is that the metamodel error can be included in the
estimators. Note that bootstrap techniques can be used 
similarly to calculate and include metamodeling error also 
for PCE-based sensitivity analysis, as demonstrated by 
\citet{Dubreuil2014}.

As shown in the various comparisons, PCE and GP give similar accuracy
(measured in terms of the $Q^2$ validation coefficient) for a given size
of the experimental design in a broad range of applications. The replication of 
the analyses with
different random designs of the same size show a smaller scatter using
GP for extremely small designs, whereas PCE becomes more stable for
medium-size designs. Selecting the best technique is in the end
problem-dependent, and it is worth comparing the two approaches using
the same experimental design, as it can be done in recent sensitivity
analysis toolboxes such as OpenTURNS  \citep{Andrianov07} 
and UQLab \citep{Marelli2014}. 

Finally it is worth mentioning that the so-called 
derivative-based
global sensitivity measures (DGSM) originally introduced by
\citet{Sobol2009} can also be computed using surrogate 
models. In
particular, polynomial chaos expansions may be used to 
compute the DGSM
analytically, as shown in \citet{SudretRESS2015}. The 
recent combination
of polynomial chaos expansions and Gaussian processes into 
{\em   PC-Kriging} \citep{SchoebiIJUQ2015} also appears 
  promising for estimating
sensitivity  indices from extremely small experimental 
designs.

\bibliographystyle{chicago}
\bibliography{biblio}

%
%
%

\end{document}